\documentclass[aps,pre,preprint,floatfix,superscriptaddress]{revtex4}
\usepackage[utf8]{inputenc}
\usepackage{amsmath}
\usepackage{graphicx}  
\usepackage{amssymb}
\usepackage{epsfig}
\usepackage{natbib}
\usepackage{grffile}

\newcommand{\BE}{\begin{equation}}
\newcommand{\EE}{\end{equation}}

\newcommand{\sech}{\mathrm{sech}} 

\usepackage[caption=false]{subfig}
\usepackage{graphicx}
\usepackage{dcolumn}
\usepackage{bm}

\def\bq{\begin{equation}}
\def\eq{\end{equation}}

\begin{document}


\title{Area-preserving maps models of gyroaveraged \\ ${\bf E}\times {\bf B}$ chaotic transport}
\author{J. D. da Fonseca}
\email{jfonseca@if.usp.br}
\affiliation{Physics Institute, University of São Paulo \\ São Paulo, SP, 5315-970, Brazil}

\author{D. del-Castillo-Negrete}
\email{delcastillod@ornl.gov}
\affiliation{Oak Ridge National Laboratory \\ Oak Ridge, TN, 37831-8071, USA}

\author{I. L. Caldas}
\email{ibere@if.usp.br}
\affiliation{Physics Institute, University of São Paulo \\ São Paulo, SP, 5315-970, Brazil}

\date{\today}

\begin{abstract}

Discrete maps have been extensively used to model 2-dimensional chaotic transport in plasmas and fluids.  Here we focus on area-preserving maps describing finite Larmor radius (FLR) effects on $E\times B$ chaotic transport in magnetized plasmas with zonal flows perturbed by electrostatic drift waves. FLR effects are included by gyro-averaging the Hamiltonians of the maps which, depending on the zonal flow profile, can have monotonic or non-monotonic frequencies. In the limit of zero Larmor radius, the monotonic frequency map reduces to the standard Chirikov-Taylor map, and, in the case of non-monotonic frequency, the map reduces to the standard nontwist map. We show that in both cases FLR leads to chaos suppression, changes in the stability of fixed points, and robustness of transport barriers. FLR effects are also responsible for changes in the phase space topology and zonal flow bifurcations. Dynamical systems methods based on recurrence time statistics are used to quantify the dependence on the Larmor radius of the threshold for the destruction of transport barriers.

\end{abstract}



\maketitle

\section{Introduction}

Important research efforts in controlled nuclear fusion are focused on the magnetic confinement of hot plasmas. 
In order to improve the confinement conditions, a better understanding of the particle transport is needed, especially in the case of 
particle drift motion. A standard approach to this problem is based on the $\vec{E}\times\vec{B}$ approximation of charged particle's guiding center's motion, e.g. 
Refs.~\cite{kleva-drake-1984,Pettini,ibere_horton_2008,del-castillo-2000}. However, in the case of fast particles, e.g. alpha particles in burning plasmas, or in the presence of 
inhomogeneous fields on the scale of the Larmor radius, it is necessary to consider finite Larmor radius (FLR) effects to correctly estimate the transport of particles \cite{Manfredi96,del-castillo-martinell2013}. 

Previous studies on the role of the Larmor radius on transport include Refs.~\cite{Annibaldi02,Manfredi96,Manfredi97} where particle transport in numerical simulations of electrostatic turbulence was analyzed
and FLR effects were shown to inhibit transport. The problem of non-diffusive chaotic transport and fractional diffusion 
in the presence of FLR effects was addressed in Ref.~\cite{gustafson}. In Refs.~\cite{del-castillo-martinell2012,del-castillo-martinell2013} dynamical systems methods were used to investigate FLR effects on chaos suppression and bifurcations in 
the phase space topology.    

In this paper we study FLR effects in the context of simplified area-preserving map models of $\vec{E}\times\vec{B}$ motion.
The maps are constructed following the standard Hamiltonian framework
for electrostatic drift motion \cite{horton-1985}, and 
FLR corrections are included by gyro-averaging the electrostatic potential  \cite{Lee87}.
The potential has an equilibrium depending on the radial coordinate and a 
time-dependent perturbation consisting of a linear superposition of drift
waves. Depending on the radial dependence of the equilibrium the maps' frequencies can have monotonic or non-monotonic profiles. 
Our main focus is the study of FLR effects on
chaos suppression, stability of fixed points, nontwist phase space topologies, and transport barriers. 
Although $\nabla \vec{B} \times \vec{B}$ drift might play a role in the
presence of FLR, here we will restrict attention to $\vec{E}\times\vec{B}$ drifts. 

In dynamical systems theory, the term ``nontwist'' is used to designate Hamiltonian systems that violate the twist or 
non-degeneracy condition. A paradigmatic example is a perturbed  system in which the 
frequency of the unperturbed part of the Hamiltonian is a non-monotonic function of the action.  
Nontwist Hamiltonian systems are found in many physical systems including 
$\mathbf{E}\times\mathbf{B}$ transport in magnetized plasmas \cite{del-castillo-2000,ibere_horton_2008}, magnetic fields with reverse shear in toroidal plasma devices \cite{Oda,Balescu98,del-castillo-1992,Firpo2009}, and 
 transport by traveling waves in shear flows  
 \cite{del-castillo-1993,Beron-Vera10,Budyansky09} among others.
The presence of nontwist transport barriers is among the most important properties of nontwist Hamiltonian systems.
By nontwist transport barriers we mean a robust region of invariant circles, also called Kolmogorov-Arnold-Moser (KAM) curves, that are very resilient  to breakup, i.e., they can survive even when the phase space is almost completely
chaotic. This property, originally discovered in Refs.~\cite{del-castillo-1993,del-castillo-negrete-etal-1996}, has also been referred to as strong KAM stability \cite{Rypina} in the context of one-and-a-half degrees-of-freedom Hamiltonian systems,
and has been the subject of several studies (see, for example, \cite{Portela07,Szezech09,Caldas,de-la-Llave-2000} and references therein). Throughout this paper use the term ``KAM curve" as a synonymous of ``invariant circle" or ``transport barrier". 

In this paper the maps correspond to Hamiltonians whose unperturbed frequencies can be  monotonic or non-monotonic functions of the action variable. The latter case is characterized by the presence of nontwist transport barriers that can be destroyed or restored by changing the value
of the Larmor radius. This property is directly related to the FLR  suppression of chaotic transport studied in Refs.~\cite{del-castillo-martinell2012, del-castillo-martinell2013}. Here we
show that chaos suppression occurs when the Larmor radius is close to specific values for which invariant circles become very resilient to breakup. 
In the case of nontwist maps, the invariant circles forming the nontwist transport barrier are the easiest to restore and the hardest to break.
One of the main goals of the present work is the numerical computation of the critical parameters for the breakup of nontwist barriers. To compute the breakup diagrams we follow the technique in 
Ref.~\cite{Altmann} which is based on the application of Slater's theorem \cite{Slater} to the recurrence properties of
orbits. As shown in Refs.~\cite{Altmann,Zou},
the recurrence properties of an orbit provide an efficient and accurate method to differentiate chaotic and non-chaotic motion. 
A recent example of the application of this technique to the standard nontwist map  was presented in Ref.~\cite{Abud}.               

The rest of the paper is organized as follows. The next section introduces a general 
area preserving map  of gyro-averaged $\vec{E}\times\vec{B}$ transport. This model is the starting point 
for the specific cases studied in the subsequent sections. 
The first case, for which the frequency has a monotonic profile, is presented in Sec.~\ref{sec:gsm}, where FLR effects on the transition to global chaos are analyzed. Section \ref{sec:gsnm} describes a case with a non-monotonic profile and discusses FLR effects on the stability of fixed points, phase space topology,  and breakup diagrams. A second nontwist map is presented in 
Sec.~\ref{sec:gqnm} where FLR effects on zonal flow bifurcations are analyzed. 
Section~VI presents the conclusions and a brief summary of the main results. 

\section{Transport Model}\label{sec:model}
The $\vec{E}\times\vec{B}$ drift velocity  of the guiding center is given by \cite{Nicholson}
\begin{equation}
   \vec{v}_{GC}  = \frac{\vec{E}\times\vec{B}}{B^{2}} \, .
   \label{eq:cg1}
\end{equation}
Using $x$ as radial coordinate, and $y$ as poloidal coordinate,  
the equations of the $\vec{E}\times\vec{B}$ drift motion, $(\dot{x}(t),\dot{y}(t))=\vec{v}_{GC}$, can be written as the Hamiltonian system
\begin{align}
 \frac{dx}{dt} =-\frac{\partial H(x,y,t)}{\partial y} \, ,   \qquad \frac{dy}{dt}=\frac{\partial H(x,y,t)}{\partial x},
   \label{eq:eqcanonicas}
\end{align}
where 
\begin{equation}
   H(x,y,t) = \frac{\phi(x,y,t)}{B_{0}}  \label{eq:ham}   \, ,
\end{equation}
$\phi$ is the electrostatic potential, and  $B_{0}$ denotes the magnitude of the toroidal magnetic field, $\vec{B} = B_{0}\hat{e}_{z}$. 
As discussed in Ref.~\cite{del-castillo-martinell2012}, finite Larmor radius (FLR) effects can be incorporated substituting the
electrostatic potential by its average over a circle around the guide center
\begin{equation}
\langle\phi\left(x,y,t\right)\rangle_{\varphi}=\frac{1}{2\pi}\int_{0}^{2\pi}\phi(x+\rho\cos\varphi,\, y+\rho\sin\varphi,\, t)\, d\varphi,\label{eq:FlrAverage}
\end{equation}
where $\rho$ is the Larmor radius. Equation~(\ref{eq:FlrAverage}) corresponds to the well-known gyro-averaging operation \cite{Lee87}. The gyro-averaged Hamiltonian can then be defined as
\begin{equation}
   \langle H(x,y,t)\rangle_{\varphi}=\frac{\langle\phi\left(x,y,t\right)\rangle_{\varphi}}{B_{0}}
   \label{eq:gyroham0-1}
\end{equation}
and the gyro-averaged equations of motion (\ref{eq:eqcanonicas}) can be written as
\begin{align}
   \frac{dy}{dt}=\frac{\partial\langle H\rangle_{\varphi}}{\partial x}, \qquad \frac{dx}{dt}=-\frac{\partial\langle H\rangle_{\varphi}}{\partial y} \, .
   \label{eq:gyrosystem}
\end{align}
Following Ref.~\cite{horton-1998}, we assume an electrostatic potential of the form
\begin{equation}
   \phi\left(x,y,t\right)=\phi_{0}(x)+A\sum_{m=-\infty}^{+\infty}\cos(ky-m\omega_{0}t),
   \label{eq:simpelepot}
\end{equation}
where $\phi_{0}(x)$ is the equilibrium potential, $A$, the amplitude of the drift waves, $k$ is the wave number, and $\omega_{0}$ is the fundamental frequency. 
Applying the gyro-average operation in Eq.~(\ref{eq:FlrAverage}) to Eq.~(\ref{eq:simpelepot}), and substituting the result in Eq.~(\ref{eq:ham}), we obtain the Hamiltonian
\begin{equation}
\langle H(x,y,t)\rangle_{\varphi}=\langle H_{0}(x)\rangle_{\varphi}+\frac{A}{B_{0}}J_{0}\left(k\rho\right)\sum_{m=-\infty}^{+\infty}\cos\left(ky-m\omega_{0}t\right),
\label{eq:gyroham}
\end{equation}
where $J_0$ is the zero-order Bessel function, and the integrable Hamiltonian $\langle H_{0}(x)\rangle_{\varphi}$ is defined as
\begin{equation}
   \langle H_0(x)\rangle_{\varphi}=\frac{\langle\phi_0\left(x\right)\rangle_{\varphi}}{B_{0}}
   \label{eq:gyroham-integrable}
\end{equation}
Using the Fourier series representation of the Dirac delta function, Eq.~(\ref{eq:gyroham}) can be rewritten as
\begin{equation}
\langle H(x,y,t)\rangle_{\varphi}=\langle H_{0}(x)\rangle_{\varphi}+\frac{2\pi A}{B_{0}}J_{0}\left(k\rho\right)\cos(ky)\sum_{m=-\infty}^{+\infty}\delta(\omega_{0}t-2\pi m).
\label{eq:gyroham2}
\end{equation}
Let  $x_n  = x(t^{-}_n)$ and $y_n  = y(t^{-}_n)$, with $t^{-}_n = \frac{2\pi n}{\omega_{0}}- \varepsilon$,  $n \in \mathbb{N}$, and $\varepsilon \rightarrow 0^{+}$.
Integrating Eq.~(\ref{eq:gyrosystem}) in the interval $(t^{-}_n,t^{-}_{n+1})$ leads to the gyro-averaged drift wave map   
\begin{align} 
   x_{n+1}&=x_n+\frac{2\pi kA}{\omega_0 B_{o}}J_{0}\left(\hat{\rho}\right)\sin(ky_n) \label{eq:dwmap-x}  \\ 
   y_{n+1}&=y_n+\frac{2\pi}{\omega_0}\Omega\left(x_{n+1}\right) \label{eq:dwmap-y}
\end{align}
where $\hat{\rho}=k\rho$, and $\Omega\left(x\right)$ corresponds to the frequency associated to the integrable Hamiltonian in Eq.~(\ref{eq:gyroham-integrable})
\begin{equation}
   \Omega\left(x\right)=\frac{d\langle H_{0}(x)\rangle_{\varphi}}{dx} \, .
   \label{eq:freq}
\end{equation}
For $\rho=0$, $\Omega\left(x\right)=-E_{r}(x)/B_0$, where $E_r(x)$ is the radial component of the electric field.
Depending on the equilibrium potential $\phi_{0}(x)$, which determines $E_{r}(x)$ and  $\Omega\left(x\right)$,
different area-preserving maps can be constructed from Eqs.~(\ref{eq:dwmap-x}) and (\ref{eq:dwmap-y}). In the next sections, we discuss three different cases.

\section{Gyroaveraged standard map}\label{sec:gsm} 
As a first simple example we assume a monotonic linear frequency profile to construct the gyro-averaged standard map (GSM). 
This map is a modified version of the standard map, also known as the Chirikov-Taylor map \cite{chirikov79,taylor}. 
To define the frequency in Eq.~(\ref{eq:dwmap-y}), we use the equilibrium potential
\begin{equation}
  \phi_{0}(x)=\alpha\frac{\left(kx\right)^{2}}{2},\label{eq:eqprof-gsm}
\end{equation}
where  $k$ is the wave number in Eq.~(\ref{eq:simpelepot}), and $\alpha$  is a free parameter. Applying the gyro-average operation to Eq.~(\ref{eq:eqprof-gsm}) yields
\begin{equation}
\langle H_0(x)\rangle_{\varphi}=\frac{\alpha}{B_{0}}\left[\frac{\left(kx\right)^{2}}{2}+\frac{\hat{\rho}^{2}}{4}\right].\label{eq:gyroham-gsm}
\end{equation}
Substituting Eq.~(\ref{eq:gyroham-gsm}) in Eq.~(\ref{eq:freq}), we get the frequency 
\begin{equation}
   \Omega\left(x\right)=\frac{\alpha k^2}{B_{0}}x,
   \label{eq:freq-gsm}
\end{equation}
which does not depend on the Larmor radius, and has a monotonic profile.
Introducing the non-dimensional variables
\begin{align}
    I=k\gamma x, \qquad \theta=ky \, ,
   \label{eq:norm-gsm}
\end{align}
and the constant
\begin{equation}
   \gamma = \frac{2\pi\alpha k^2}{\omega_0 B_{0}} \, ,
   \label{eq:gamma}
\end{equation}
we get, from Eq.~(\ref{eq:dwmap-x}) and Eq.~(\ref{eq:dwmap-y}), the gyro-averaged standard map (GSM)
\begin{align}
  I_{n+1}= & I_{n}+K_{ef}\left(\hat{\rho}\right)\sin\theta_{n}\label{eq:mapI-norm-2}\\
  \theta_{n+1}= & \theta_{n}+I_{n+1},\quad mod\quad2\pi\label{eq:mapTheta-norm-2}
\end{align}
where
\begin{equation}
   K_{ef}=KJ_{0}\left(\hat{\rho}\right)
\end{equation}
is the effective perturbation parameter and $K=\gamma^2A /\alpha$ is the perturbation parameter. For $\hat{\rho}=0$, $K_{ef}=K$, and the GSM reduces to the standard map. 

The phase space of the GSM, as in the case of any
area-preserving map, consists of periodic, quasiperiodic, and chaotic
orbits. Quasiperiodic orbits cover densely invariant curves. The invariant curves around the elliptic fixed points form island chains, and the invariant curves that wind around the entire domain of the angle variable form invariant circles \cite{Meiss92}. The presence of KAM curves is of special interest because they preclude transport in the direction of the radial coordinate $x$, 
keeping chaotic orbits confined to specific regions of phase space.

\begin{figure}
\includegraphics[scale=0.20]{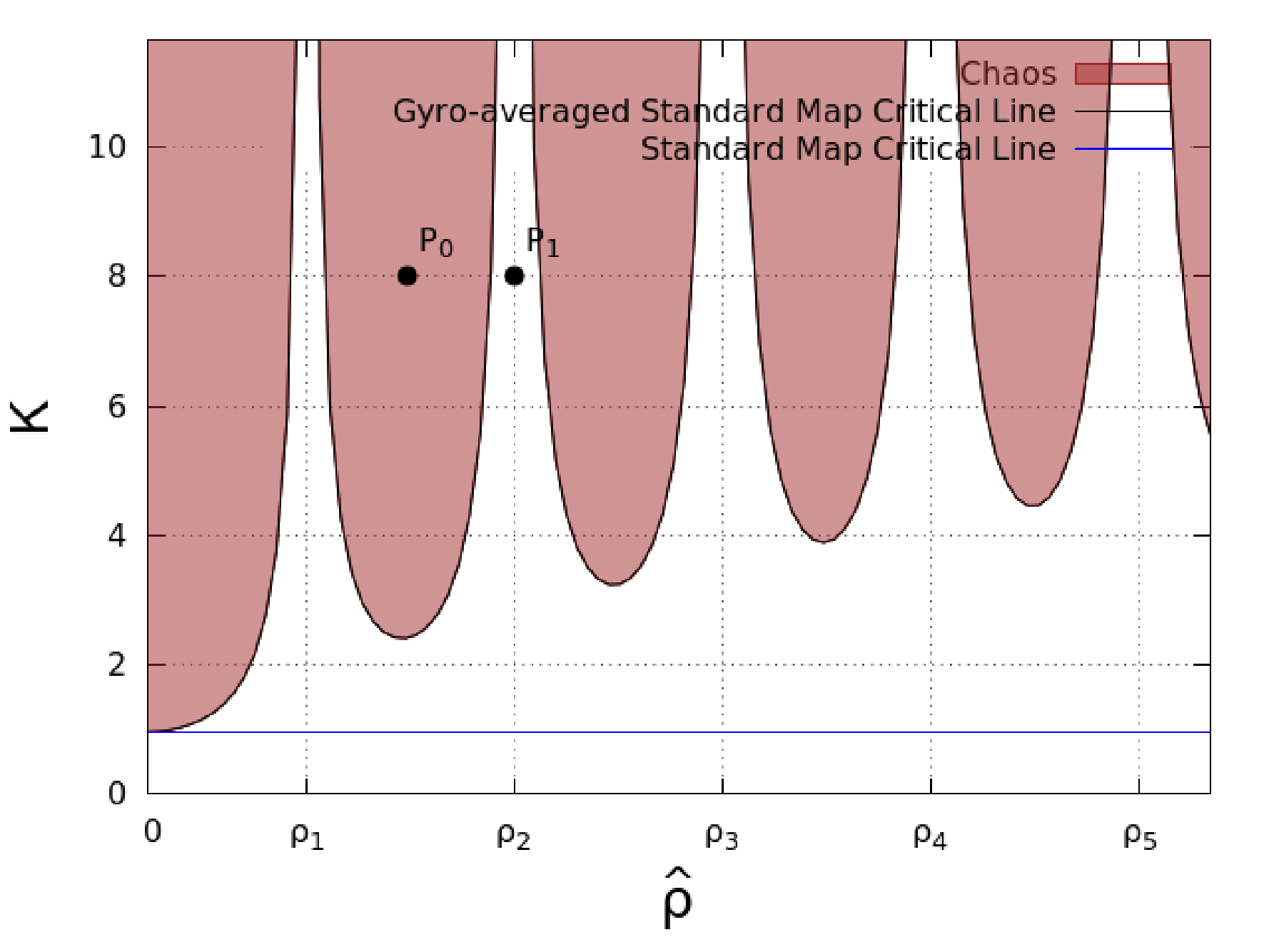}
\caption{Critical lines for the standard map (blue horizontal line) and for the gyro-averaged standard map (black curves).
Inside the shaded regions there is global chaos. (Color online)}
\label{fig_threshold_twist_map}
\end{figure}

According to Greene's residue method \cite{Greene79}, the transition to global chaos in the standard map occurs when the absolute value of 
the perturbation parameter is equal to, or greater than, the critical value $K_{c}\simeq0.9716$. That is, for $K \geq K_{c}$, all KAM curves are broken, and chaotic orbits can spread over all phase space
(except in regions occupied by isolated islands).
Larmor radius effects on the transition to global chaos in the GSM can be analyzed by defining the critical line dividing the $K-\hat{\rho}$ parameter space in two regions: one for which the phase space contains at least one KAM curve and another for which transition to global chaos has occurred. In the standard map,
there is no dependence on $\hat{\rho}$, which means that the critical line is just a horizontal line defined at $K=K_{c}$, as indicated in Fig.~\ref{fig_threshold_twist_map}. 
However, for the GSM, the critical line is determined by the condition $\left|K_{ef}\right|=K_{c}$ which implies
\begin{equation}
K=\frac{K_{c}}{\left|\, J_{0}(\hat{\rho})\right|} 
\label{cr_line-gsm}
\end{equation}
As shown in Fig.~\ref{fig_threshold_twist_map}, even for high values of the perturbation
parameter ($ K \gg K_{c} $) there are an infinite number of regions of Larmor radius values for which KAM curves can be restored and chaos is suppressed. The gyro-averaging operation ``breaks'' the critical line at the zeros of the zero-order Bessel function. 
In particular, near the zeros,  the critical perturbation goes to infinity and the transition to global chaos
cannot occur. 
Writing $\hat{\rho}=\rho_i+\delta \hat{\rho}$, where $\rho_i$ is the $i$-th zero of $J_0$ and $\delta \hat{\rho} \ll 1$, 
and expanding in Taylor series, the divergence
of the critical  perturbation near the zeros of $J_0$ can be approximated as
\bq
K\approx \frac{K_c/|J_1(\rho_i)}{|\hat{\rho}-\rho_i|} \, ,
\eq
where $J_1$ denotes the Bessel function of first order. The first five positive zeros, indicated  in Fig.~ \ref{fig_threshold_twist_map},
are approximately $\rho_1=2.40$, $\rho_2=5.52$, $\rho_3=8.65$, $\rho_4=11.79$, and $\rho_5=14.93$. 

\begin{figure}
\includegraphics[scale=0.35]{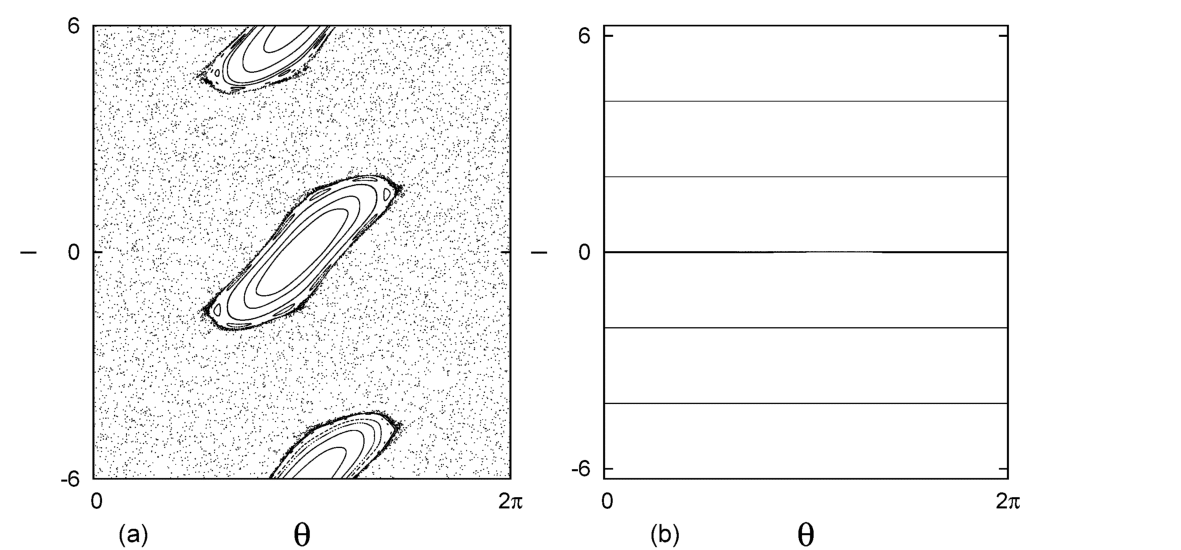}
\caption{(a) GSM Poincaré section for $\hat{\rho}=3.9$ and $K=8.0$ (point $P_0$ in figure \ref{fig_threshold_twist_map}): no KAM curves are observed.
(b) GSM Poincaré section for $\hat{\rho}=5.52$ and $K=8.0$ (point $P_1$ in figure \ref{fig_threshold_twist_map}): KAM curves are restored.}
\label{fig_p1p2}
\end{figure}

The effect of chaos suppression is illustrated by Figs.~\ref{fig_p1p2} (a)-(b). Figure \ref{fig_p1p2}-(a) shows a GSM 
Poincaré section for $\hat{\rho}=3.9$ and $K=8.0$,
which correspond to the point $P_0$ in Fig.~\ref{fig_threshold_twist_map}. No KAM curves are observed in the
Poincaré section as $P_0$ belongs to the region of global chaos. Keeping the same value of $K$ and changing 
the value of $\hat{\rho}$  to $\rho_2$ (point $P_1$ in Fig.~\ref{fig_threshold_twist_map}) 
eliminates all the chaotic orbits (see figure \ref{fig_p1p2}(b)).

\section{Gyroaveraged standard nontwist map}\label{sec:gsnm} 
As a second example of the gyro-averaged drift wave map, we introduce the gyroaveraged standard nontwist map (GSNM),
which corresponds to a non-monotonic radial electric field. In this case the equilibrium potential is  
\begin{equation}
\phi_{0}(x)=\alpha\left[\left(\frac{x}{L}\right)-\frac{1}{3}\left(\frac{x}{L}\right)^{3}\right],\label{eq:eqprof-gsnm}
\end{equation}
where $\alpha$ and $L$ are dimensional constants. The gyro-average of Eq.~(\ref{eq:eqprof-gsnm}) gives
\begin{equation}
\langle H_{0}(x)\rangle_{\varphi}=\frac{\alpha}{B_{0}}\left[\left(\frac{x}{L}\right)\left(1-\frac{\rho^{2}}{2L^{2}}\right)-\frac{1}{3}\left(\frac{x}{L}\right)^{3}\right],
\label{eq:gyroeqham-gnsm}
\end{equation}
and from Eq.~(\ref{eq:freq}) we get the non-monotonic parabolic frequency profile,
\begin{equation}
  \Omega\left(x\right)=\frac{\alpha}{B_{0}L}\left\{ \left[1-\left(\frac{\rho}{\sqrt{2}L}\right)^{2}\right]-\left(\frac{x}{L}\right)^{2}\right\} \, ,
  \label{eq:freq-gsnm}
\end{equation}
which has a maximum at $x=0$. 
For $\rho=0$, the zeros of $\Omega\left(x\right)$  (and also of the $E_r$ profile) are located at $\pm L$, where $L$ is 
the characteristic length of the frequency profile. Substituting Eq.~(\ref{eq:freq-gsnm}) into Eq.~(\ref{eq:dwmap-y}),
we obtain
\begin{align}
   x_{n+1}= & x_{n}+\frac{2\pi}{\omega_{0}B_{0}}AkJ_{0}\left(k\rho\right)\sin\left(ky_{n}\right)\label{eq:gsnm-x}\\
   y_{n+1}= & y_{n}+\frac{2\pi}{\omega_{0}B_{0}}\frac{\alpha}{L}\left\{ \left[1-\left(\frac{\rho}{\sqrt{2}L}\right)^{2}\right]-\left(\frac{x_{n+1}}{L}\right)^{2}\right\}\label{eq:gsnm-y} .
\end{align}
Introducing the dimensionless variables:
\begin{align}
    I=-\frac{x}{L}, \qquad \theta=\frac{ky}{2\pi},
   \label{eq:norm-gsnm}
\end{align}
we get the gyro-averaged standard nontwist map (GSNM):
\begin{align}
   I_{n+1}= & I_{n}-bJ_{0}\left(\hat{\rho}\right)\sin\left(2\pi\theta_{n}\right)\label{eq:gsnm-I}\\
   \theta_{n+1}= & \theta_{n}+a\left[\left(1-\frac{\bar{\rho}^{2}}{2}\right)-I_{n+1}^{2}\right],\quad {\rm mod}\,1\label{eq:gsnm-theta}
\end{align}
where:
\begin{eqnarray}
 a =& \frac{\alpha k}{\omega_0 B_{0}L} \qquad b=& \frac{2\pi aA}{\alpha} \label{eq:gsnm-paramab}\\
\bar{\rho} =&\frac{\rho}{L} \qquad \hat{\rho}= k\rho \label{eq:gsnm-paramrho}
\label{eq:gsnm-param}
\end{eqnarray}
are four dimensionless parameters.
We refer to $b$  as the perturbation parameter, which is proportional to the amplitude $A$ of the drift waves.
Like in the GSM case, we can also define an effective perturbation parameter
\begin{equation}
   b_{ef} = bJ_{0}\left(\hat{\rho}\right)\, . \label{eq:gsnm-efpert}
\end{equation}
As indicated in Eq.~(\ref{eq:gsnm-paramrho}), $\bar{\rho}$, and
$\hat{\rho}$
correspond to the Larmor radius normalized using two different length scales:
the  characteristic length of the frequency profile $L$ and the wavelength $\lambda=\frac{2\pi}{k}$ respectively

	\subsection{Fixed Points}\label{subsec:gsnm-fixedpoints}

\begin{figure}
\begin{centering}
\includegraphics[scale=0.325]{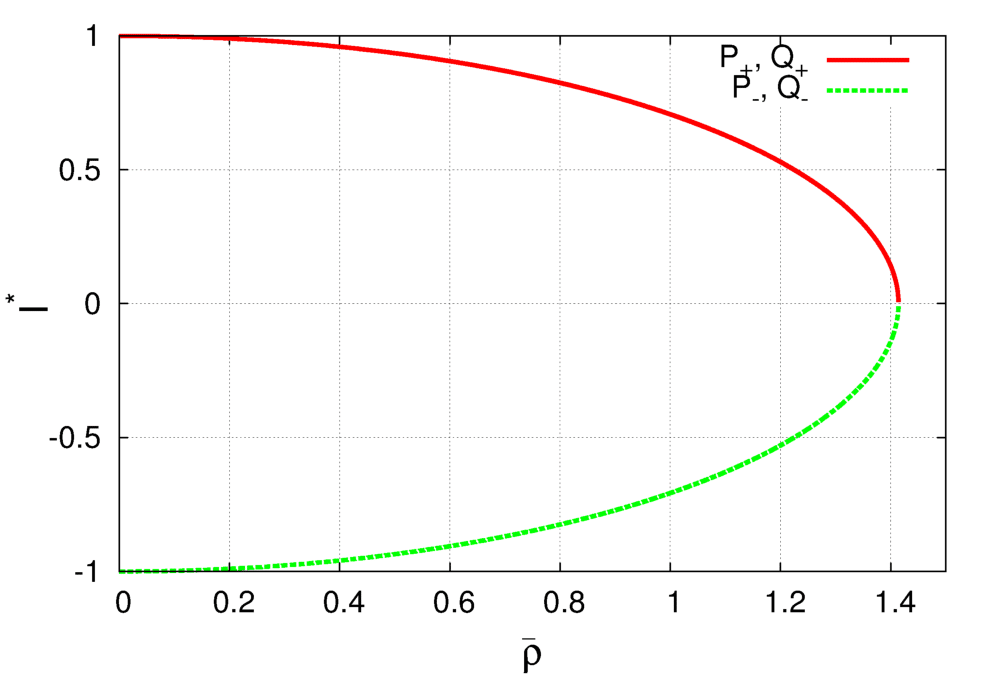}
\par\end{centering}
\caption{Coordinate $I$ of fixed points for m=0 and increasing $\bar{\rho}$. The fixed points collide at  $\bar{\rho}=\sqrt{2}$ and do not exist for $\bar{\rho}>\sqrt{2}$. (Color online)}
\label{IcoordGNSM}
\end{figure}

According to Eqs.~(\ref{eq:gsnm-I})-(\ref{eq:gsnm-theta}), 
the fixed points of the GSNM,
$I^{*}=I_{n+1}=I_{n}$ and $\theta^{*}=\theta_{n+1}=\theta_{n}$,  satisfy
\begin{align}
   0= & bJ_{0}\left(\hat{\rho}\right)\sin(2\pi\theta^{*})  \label{eq:gsnm-fpI}\\
   m= & a\left[\left(1-\frac{\bar{\rho}^{2}}{2}\right)-I^{*}{}^{2}\right]   \label{eq:eq:gsnm-fpTheta}
\end{align}
where $m$ is an integer number. For each  $m\in\mathbb{\mathbb{Z}}$, $a\neq0$, $bJ_{0}\left(\hat{\rho}\right)\neq0$,
and $\bar{\rho}\leq\sqrt{2\left(1-\frac{m}{a}\right)}$, there are
four fixed points: 
\begin{align}
   P_{\pm}= \left(0,\pm I_{*}\left(\bar{\rho}\right)\right) \label{eq:fp-P}\\
   Q_{\pm}= \left(\frac{1}{2},\pm I_{*}\left(\bar{\rho}\right)\right) \label{eq:fp-Q}
\end{align}
where
\begin{equation}
   I_{*}\left(\bar{\rho}\right)=\frac{1}{\sqrt{2}}\sqrt{2\left(1-\frac{m}{a}\right)-\bar{\rho}^{2}} \, .
   \label{eq:gsnm-fpHeight}
\end{equation}
Note that the $\theta$ coordinate of $P_{\pm}$ and $Q_{\pm}$ does not depend on any parameters. 
If $\bar{\rho}=\sqrt{2\left(1-\frac{m}{a}\right)}$, the pair of points $\left\{P_{+},P_{-}\right\}$ 
collide at $\left(0,0\right)$, and  $\left\{Q_{+},Q_{-}\right\}$ collide at $\left(\frac{1}{2},0\right)$. Figure \ref{IcoordGNSM} shows the $I$-axis position
of both pairs for $m=0$ and increasing $\bar{\rho}$. The collision occurs for $\bar{\rho}=\sqrt{2}$. For higher values of $\bar{\rho}$, the fixed points
don't exist.

\begin{figure}
\begin{center}
\includegraphics[scale=0.45]{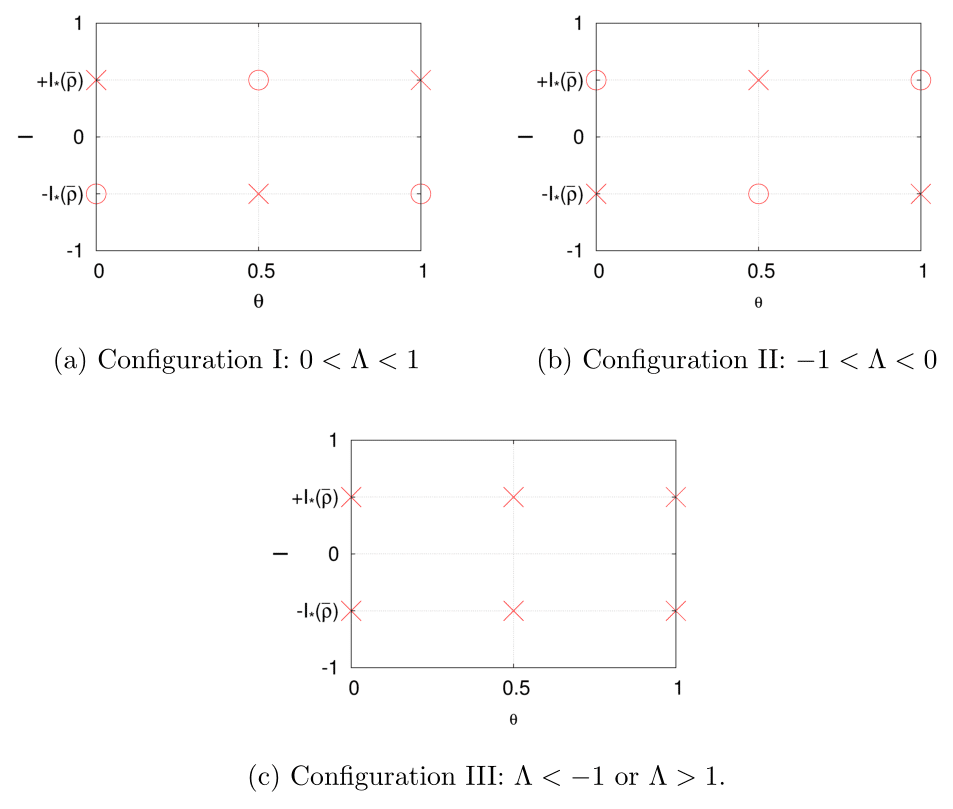}
\caption{Stability of the fixed points $P_{\pm}$ and $Q_{\pm}$with
$m=0$. (Color online)}
\label{fig:configGNSM}
\end{center}
\end{figure}

The stability of a $k$-periodic orbit of a map $M$ is determined by the residue
\begin{equation}
   R=\frac{1}{4}\left[2-Tr\left(\prod_{i=0}^{k-1}J\left(\vec{x}_{i}\right)\right)\right]
  \label{eq:res}
\end{equation}
where $J$ is the Jacobian matrix of $M$, evaluated at 
the $k$-periodic orbit $\left\{\vec{x}_{i}\right\} _{i=0}^{k-1}$.
If $0<R<1$, the periodic orbit is elliptic (stable); if $R<0$ or $R>1$,
it is hyperbolic (unstable); and it is parabolic for $R=0$ or $R=1$.
Applying formula (\ref{eq:res}) to the GNSM fixed points in Eqs.~(\ref{eq:fp-P})-(\ref{eq:fp-Q}) we get
\begin{equation}
   R\left(P_{\pm}\right)=R\left(Q_{\mp}\right)=\mp \pi abJ_{0}\left(\hat{\rho}\right) I_{*}\left(\bar{\rho}\right) \, .
   \label{eq:res-PQ}
\end{equation}
The stability of the fixed points $\left\{P_{\pm},Q_{\pm}\right\}$ with $m=0$ can be analyzed using the 
parameter $\Lambda\left(a,b,\hat{\rho},\bar{\rho}\right)=
R\left(P_{-}\right)$.
As shown in Fig.~\ref{fig:configGNSM}, depending on the value of $\Lambda$, there are three possible configurations.
The symbol ``x'' denotes an hyperbolic point, and ``o'' an elliptic point. The points have their stability inverted when $-1<\Lambda<0$ (configuration \emph{II}), as indicated in Fig.~\ref{fig:configGNSM}-(b).
All the fixed points are hyperbolic in configuration \emph{III}, Fig.~\ref{fig:configGNSM}-(c), which occurs for $\Lambda<-1$ or $\Lambda>1$.

\subsection{Separatrix reconnection}
\begin{figure}
\begin{center}
\includegraphics[scale=0.4]{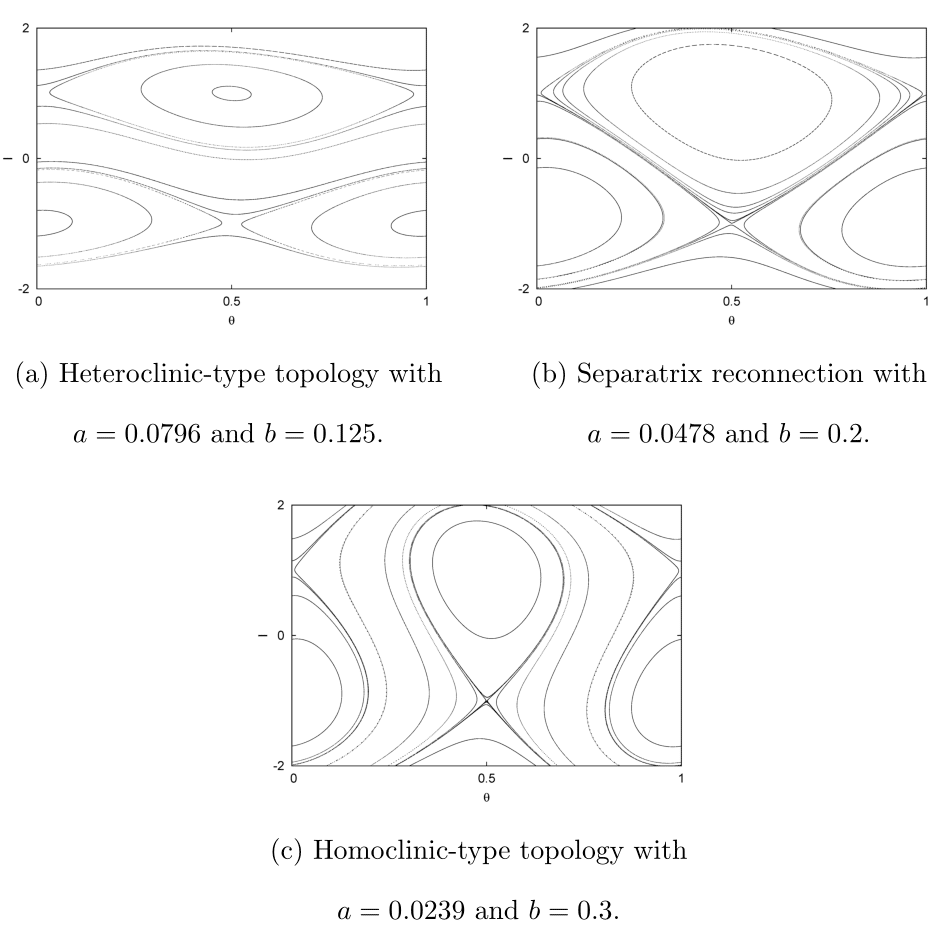}
\caption{GSNM phase space topologies associated with the fixed points $P_{\pm}$ and $Q_{\pm}$ with $m=0$. In all figures, the Larmor radius is such that $\hat{\rho}=0.05$ and $\bar{\rho}=0.01$. }
\label{fig:topologies}\end{center}
\end{figure}

The location and stability of the fixed points determine the different phase space topologies of the GSNM.
These topologies, illustrated in Fig.~\ref{fig:topologies}, are  characteristic of nontwist maps and
are called heteroclinic, separatrix reconnection, 
and  homoclinic \cite{del-castillo-1993}. Since the Larmor radius changes the stability of the fixed points, it is expected that it will also 
change the topology. Figures \ref{fig:topologies}(a)-(c) show the heteroclinic-type,
separatrix reconnection, and the homoclinic-type topologies in Poincaré sections of the GSNM for 
different values of $a$ and $b$, and fixed values of $\hat{\rho}$ and $\bar{\rho}$. 

To determine the condition for separatrix reconnection associated to the fixed points $P_{\pm}$ and $Q_{\pm}$ with $m=0$,
we follow Refs.~\cite{del-castillo-negrete-etal-1996,del-castillo-1993} and approximate the GSNM in the vicinity of the fixed points 
by the Hamiltonian
\begin{equation}
   H\left(I,\theta\right)=a\left[I\left(1-\frac{\bar{\rho}^{2}}{2}\right)-\frac{I^{3}}{3}\right]-\frac{b}{2\pi}J_{0}\left(\hat{\rho}\right)\cos\left(2\pi\theta\right) \, . \label{eq:gsnm-aproxham}
\end{equation}
For $0<\Lambda<1$ (configuration I), separatrix reconnection occurs when $H\left(P_{+}\right)=H\left(Q_{-}\right)$.
If $-1<\Lambda<0$ (configuration II), reconnection is observed when $H\left(P_{-}\right)=H\left(Q_{+}\right)$. 
Combining these two conditions, we obtain the reconnection line
\begin{align}
    a=\frac{3}{4\pi}\sigma\left(\hat{\rho},\bar{\rho}\right)b
   \label{eq:reconline}
\end{align}
with slope:
\begin{equation}
   \sigma\left(\hat{\rho},\bar{\rho}\right)=\frac{\left|J_{0}\left(\hat{\rho}\right)\right|}{\left(1-\frac{\bar{\rho}^{2}}{2}\right)^{\frac{3}{2}}}
   \label{eq:sigma}
\end{equation}

\begin{figure}[!h]
\begin{centering}
\includegraphics[scale=0.17]{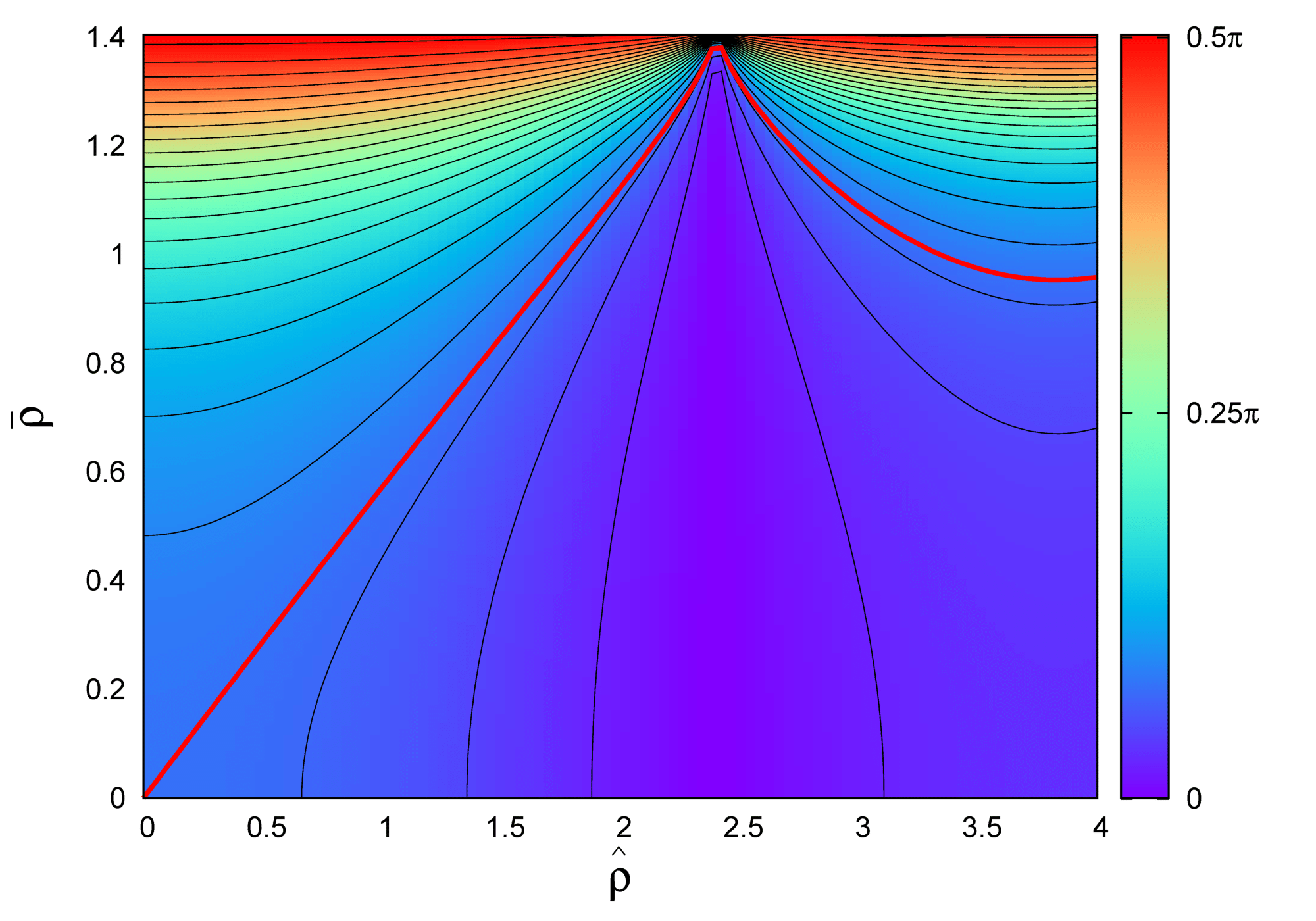}
\par\end{centering}
\caption{Isolines of $\arctan\left[\frac{3}{4\pi}\sigma\left(\hat{\rho},\bar{\rho}\right)\right]$: varying
$\hat{\rho}$ and $\bar{\rho}$ along isolines does not change the topology of the phase space. For high (red) values
of $\sigma$, homoclinic-type is observed
topology for a fixed $a$ and  almost every value of $b$. For low (blue)
values of $\sigma$, heteroclinic-type topology is observed for a fixed $b$ and almost every value of $a$. (Color online)}
\label{fig:isosigma}
\end{figure}

The reconnection line divides
the $a$-$b$ parameter space  in two regions: 
one with heteroclinic-type topology ($\frac{a}{b} > \frac{3}{4\pi}\sigma\left(\hat{\rho},\bar{\rho}\right)$) and another with homoclinic-type
topology ($\frac{a}{b} < \frac{3}{4\pi}\sigma\left(\hat{\rho},\bar{\rho}\right)$). The slope of the reconnection line 
is defined by the angle $\arctan\left[\frac{3}{4\pi}\sigma\left(\hat{\rho},\bar{\rho}\right)\right]$. 
Figure \ref{fig:isosigma} shows isolines of the angle
$\arctan\left[ \frac{3}{4\pi}\sigma\left(\hat{\rho},\bar{\rho}\right)\right]$.
Keeping $a$ and $b$ fixed, the topology of the phase space remains
unchanged if the parameters $\hat{\rho}$ and $\bar{\rho}$ vary over
an isoline. As an example, for $a$ and $b$ such that $a=\frac{3}{4\pi}b$ (which corresponds to the reconnection condition in the
standard nontwist map \cite{del-castillo-1993})  there is reconnection
for any values of $\hat{\rho}$ and $\bar{\rho}$ over the red isoline in Fig.~\ref{fig:isosigma}. 
The red isoline crosses the  point $\hat{\rho}=\bar{\rho}=0$ where the
GNSM reduces to the standard nontwist map. For high (red) values of $\sigma$, the slope of the reconnection
line approaches $\frac{\pi}{2}$, and the phase space is characterized
by the homoclinic-type topology for a fixed $a$ and almost every value of the parameter
$b$. For low (blue) values of $\sigma$, the slope of the reconnection
line tends to $0$, and the phase space is characterized by the
heteroclinic-type topology for a fixed $b$ and almost every value of the parameter $a$.

	\subsection{Nontwist transport barrier}\label{sec:ntb}

The transition to global chaos in the GSNM  corresponds to  the destruction of the nontwist transport barrier (NTB)
which is a robust non-chaotic region of KAM curves dividing the phase space. 
The NTB is robust in the sense that their KAM curves are typically more resilient to perturbation than other KAM curves in the system.

\begin{figure}
\includegraphics[scale=0.325]{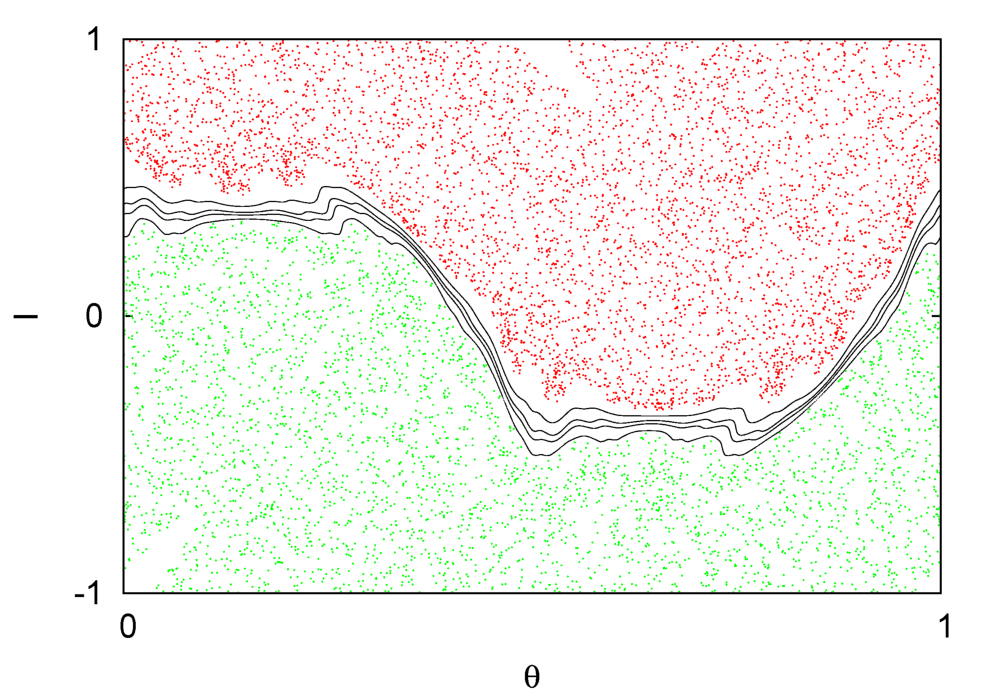}
\caption{Nontwist transport barrier in the GSNM: robust region of KAM curves (black curves) splitting the phase space in two chaotic regions,
indicated by the red and green orbits. (Color online)}
\label{fig:gsnm-barrier}
\end{figure}

Consider the  $3/2$-degrees-of-freedom Hamiltonian system
\begin{equation}
   H=H_{0}\left(I\right)+\epsilon V\left(I,\theta,t\right) \label{eq:H3/2}
\end{equation}
where $I$ and $\theta$ are the action-angle variables of the integrable system, $H_{0}$,
and the parameter $\epsilon$ controls the strength of the time-periodic perturbation $V$.  In this case, the twist (non-degeneracy) condition is 
\begin{equation}
   \frac{d\Omega}{dI}=\frac{d^{2}H_{0}}{dI^{2}}\neq0, \label{eq:twist-cond}
\end{equation}
for all $I$. If there is at least one critical point for which $d\Omega/dI=0$,
the non-degeneracy condition is violated the Hamiltonian in Eq.~(\ref{eq:H3/2}) is nontwist. 
One of the main properties of these systems is the presence of robust NTBs located at the  critical points (maxima or minima) of the 
frequency profile even in the case when most of the phase space is chaotic \cite{del-castillo-1993}.
As discussed before, the GSNM can be obtained 
by integrating the equations of motion of the $3/2$-degrees of freedom Hamiltonian in Eq.~(\ref{eq:gyroham}). This is a nontwist Hamiltonian because the frequency in  Eq.~(\ref{eq:freq-gsnm}) has a critical point at $x=0$. 
Figure \ref{fig:gsnm-barrier} shows a  Poincaré section of the GSNM with 
$a=0.354$, $\bar{\rho}=0$, and $bJ_{0}\left(\hat{\rho}\right)=0.8$.
As expected, the phase space is characterized predominantly by chaotic orbits (green and red regions), but it
has a robust NTB consisting of ``belt'' of KAM curves (black curves). 
	\subsection{Breakup Diagrams}

Here we study the transition to global chaos in the GSNM, i.e., the destruction of the robust KAM curves that constitute the NTB.
It is important to remark that
the transition to global chaos in nontwist systems is still an open problem. However, a possible
approach to this problem consists of estimating 
the parameter values, or the critical thresholds, that determine the breakup of the shearless KAM curve. 
In the absence of a perturbation,  the shearless
KAM curve is located where the twist condition is violated, and the parameter values for the  destruction of the NTB can be estimated from the parameter values that determine the breakup of the shearless curve.   

A standard way to determine the parameter values for the destruction of the shearless curve is based on the indicator point (IP) method 
proposed in Ref.\cite{shinohara} and used is several works including 
Refs.~\cite{Wurm,Budyansky09,del-castillo-martinell2013}. 
This method consists on finding the parameter values for which the iterations of the indicator point, which defines 
the indicator point orbit (IPO), are chaotic. If the IPO is quasiperiodic, it traces a shearless KAM curve. 
When the IPO is chaotic, the shearless curve is broken. In nontwist maps with special symmetries, like the ones studied here, 
the indicator points can be easily found  by computing the fixed points 
of the reversing symmetry group transformations. Further details can be found in Ref.~\cite{Petrisor}.  
Following a procedure similar to the one in Ref.~\cite{Wurm}, we obtained the following IPs for 
the gyro-averaged standard nontwist map
\begin{equation}
   z_{0}^{1} = \left(\frac{1}{4},\frac{bJ_{0}\left(\hat{\rho}\right)}{2}\right) \quad z_{0}^{2}=\left(\frac{3}{4},-\frac{bJ_{0}\left(\hat{\rho}\right)}{2}\right)\\
   \label{eq:ipO}
\end{equation}
\begin{equation}
z_{1}^{1}=\left(\frac{a}{2}\left[1-\frac{\bar{\rho}^{2}}{2}\right]+\frac{1}{4}\quad {\rm mod}\quad1,\quad  0 \right) \qquad z_{1}^{2}= \left(\frac{a}{2}\left[1-\frac{\bar{\rho}^{2}}{2}\right]+\frac{3}{4}\quad {\rm mod}\quad1,0\right)
\label{eq:ip1}
\end{equation}

Once the IPs are found, the next step is to obtain the critical thresholds using
an appropriate criterion to determine if the IPO
is chaotic. Our criterion is based on the technique proposed in \cite{Altmann}, which considers
the recurrence properties of the IPO in conjunction with 
Slater's Theorem \cite{Slater}. Recurrence properties and their relation to the Slater's Theorem were also discussed in \cite{Zou}. In Ref.~\cite{Abud}, the technique described in Ref.~\cite{Altmann}
was used to study the breakup of the shearless curve in the nontwist standard map. 
The Slater's Theorem states that, 
for any quasiperiodic motion on the circle, there are at most three different recurrences, or return times,
in any connected interval. Although this theorem was originally formulated for
circle maps, it can also be applied to two-dimensional systems, once the dynamics on a KAM curve is mapped to a quasiperiodic rotation on a circle \cite{Altmann}. 
The recurrence time is defined as the number of iterations that an orbit takes to return to a neighborhood
of a point. Our procedure to characterize the dynamics of an indicator point orbit of the GSNM is the following:  
with any one of the indicator points as the initial condition, we compute the orbit for $N$ iterations and choose the point of the orbit which has 
the maximum number of different recurrence times.  If there are more than three different recurrence times,
we conclude that the orbit is chaotic and the corresponding KAM curve is destroyed. 
More precisely, the procedure consists of the following steps:
\begin{itemize}
\item Compute the orbit $O = \{u_k\}_{k=0}^{N}$ with initial condition $u_0 = z_{1}^{1}$, where $N$ is the number of iterations.
\item Construct the recurrence matrix \footnote{Here we borrow the definition presented in Ref. \cite{Zou} of the
binary matrix $R$. Its graphical representation, called recurrence plot, can be used to analyze the recurrence properties of any dynamical system.}:
\begin{equation}
   R_{ij} = \Theta(\epsilon - \left\Vert u_i - u_j\right\Vert)
\end{equation}
where $i,j\in\left\{0,...,N\right\}$, $u_i,u_j\in O$, $\Theta$ is  the Heaviside function, and $\epsilon$ is a parameter defining the size of the neighborhood.
If the distance between $u_i$ and $u_j$, given by the norm $\left\Vert u_i - u_j \right\Vert$, is less than $\epsilon$, $R_{ij}=1$;
otherwise, $R_{ij} = 0$. 
\item Define the recurrence time as $\tau_{ij}= \left|i-j\right|$ for $i\neq j$ and $R_{ij} = 1$. That is, the recurrence time is $\tau_{ij}$ if 
the orbit  crosses the neighborhood of $u_j$ at the $i$-th iteration. For each point $u_j$, 
compute the set of different recurrences $S^{(j)}_R = \cup_{i=0,i\neq j}^{N} \left\{ \tau_{ij}\right\}$. 
\item Determine the maximum number of different recurrence times:
\begin{equation}
    n_R = max\left\{n\left(S^{(j)}_R\right)\right\}_{j=0}^{N} 
\end{equation}
where $n$ is the number of recurrence times belonging to $S^{(j)}_R$, and use the Slater's Theorem to conclude that the orbit $O$ is chaotic if $n_{R}  > 3$. 
\end{itemize}
For $n_{R}\leq3$, the characterization of the dynamics using the Slater's Theorem  is inconclusive because 
the orbit might be periodic, quasiperiodic or even chaotic. In this case, 
the number of iterations, $N$, can be increased to explore if $n_R$ increases beyond $3$. If the condition $n_{R}\leq3$
still persists and no change in $n_{R}$ is observed, the orbit is deemed periodic if $n_R=1$ and quasiperiodic if 
$n_R=3$. The case $n_R=2$ requires a more careful analysis with higher values $N$.  Increasing $N$, $n_R$ stabilizes in $3$, indicating quasiperiodic dynamics,
or assumes values greater than $3$, which is the case for chaotic dynamics. 
Although rigorous criteria for the case $n_{R}\leq3$ are lacking,
the numerical results in Refs. \cite{Altmann, Zou} support the use of recurrences 
in conjunction with Slater's theorem as a computationally efficient diagnostic to identify chaotic orbits. 

\begin{figure}
\begin{center}
\includegraphics[scale=0.35]{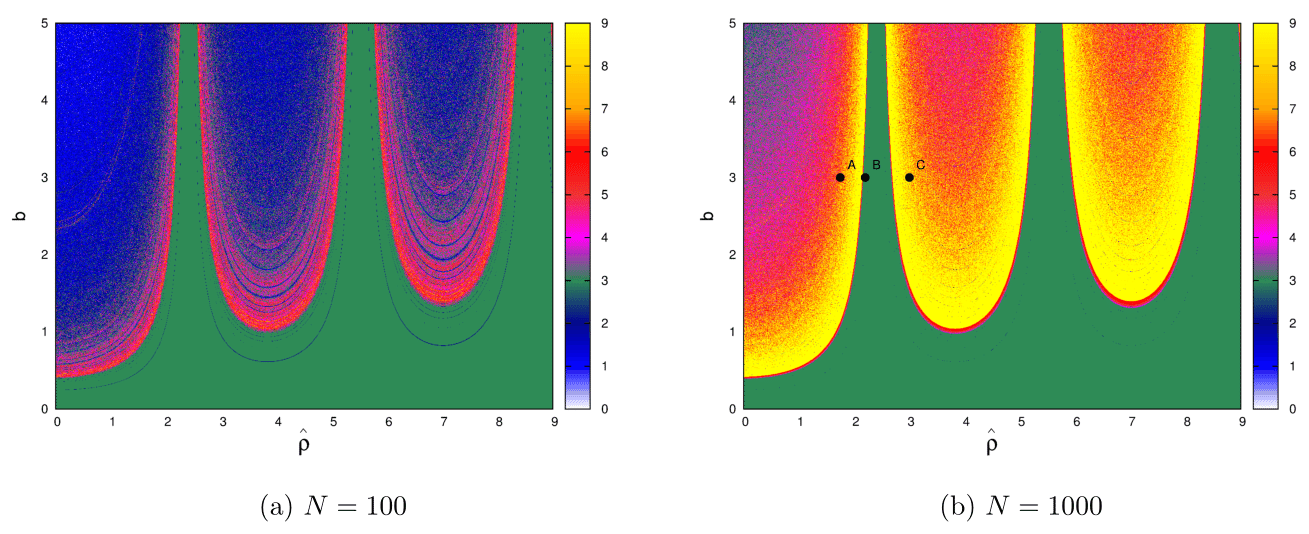}
\caption{GSNM breakup diagrams for $\hat{\rho}$ versus $b$,
         with $\bar{\rho}=0$, and $a=0.1$. For each point $(\hat{\rho},b)$ in the diagram, the IPO and the correspondent $n_R$
         are computed with $\epsilon=0.1$. The color pallet indicates the value of $n_R$. The breakup of the shearless curve can be detected
         in the points where $n_R > 3$. The IPO is quasiperiodic in the green region ($n_R = 3$), which corresponds to low effective
         perturbation $b_{ef}$, i.e., low $b$ or $\hat{\rho}$ near a zero of $J_{0}$. (Color online)}
          \label{fig:bVSrhohatN100}  
\end{center}
\end{figure}

\begin{figure}
\begin{center}
\includegraphics[scale=0.4]{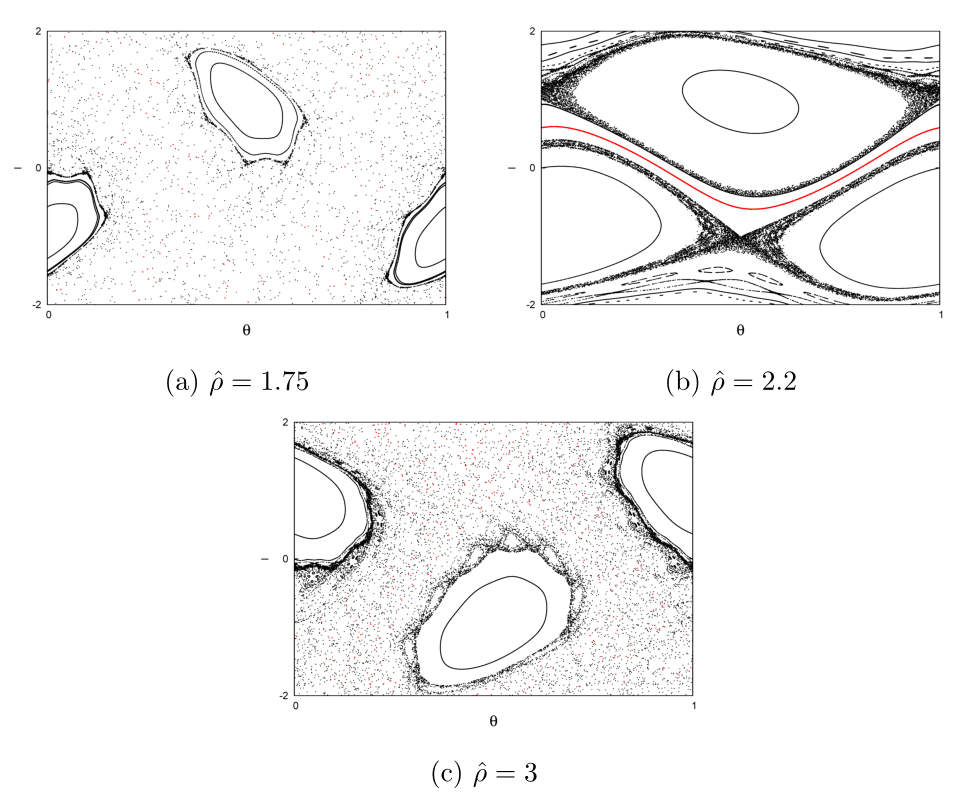}
\caption{Restoration of  shearless KAM curve (identified through a quasiperiodic IPO shown in red) 
for $\hat{\rho}$ values near the zeros of $J_0$ and  $a=0.1$, $b=3$, and  $\bar{\rho}=0$.
Panels (a), (b) and (c) correspond to parameter values  A, B, and C indicated in Fig.~\ref{fig:bVSrhohatN100}. 
As expected, for $\hat{\rho}=2.2$, which is near a zero of the Bessel function, the effective perturbation parameter is small and the shearless curve is restored. (Color online)}
 \label{fig:brhohat}
\end{center}
\end{figure}

We now apply this diagnostic to compute breakup diagrams for the GSNM showing regions in parameter
space where the shearless curve is broken. The diagrams were constructed by 
computing the recurrences of IPOs for $\epsilon=0.1$, and two different number of iterations,  $N=100$ and $N=1000$. 
Since the GSNM has four free parameters only two parameters were varied at a time while the others were kept constant. Our first example is presented in Fig.~\ref{fig:bVSrhohatN100} where  we fixed $a=0.1$,  $\bar{\rho}=0$ (which corresponds to the limit $\rho \ll L$), and 
plotted the values of $n_r$ on a   $1000 \times 1000$ grid in the $(\hat{\rho},b)$ space. 
The main difference between $N=100$ and $N=1000$ is observed in points with $n_R = 2$, but  even for a relatively small number of iterations it is possible to identify the breakup of the 
shearless curve in domains where $n_R > 3$. 
These domains have points plotted with magenta, red, and yellow colors. 
Green points ($n_R = 3$) are concentrated in domains with low effective perturbation $b_{ef}$, 
which means small $b$ or  $\hat{\rho}$ close to a zero
of the zero-order Bessel function. For parameter values outside the green region (e.g., points $A$ and $C$ in Fig.~\ref{fig:bVSrhohatN100}) the IPO is chaotic, and, 
as shown in Fig. \ref{fig:brhohat}, the shearless curve
is destroyed.  For $\hat{\rho}$ values near a zero of $J_{0}$  
(e.g., point $B$ of Fig.~\ref{fig:bVSrhohatN100}),
the shearless KAM curve is restored as shown by the red orbit in Fig.~\ref{fig:brhohat}. 

\begin{figure}
\begin{center}
\includegraphics[scale=0.35]{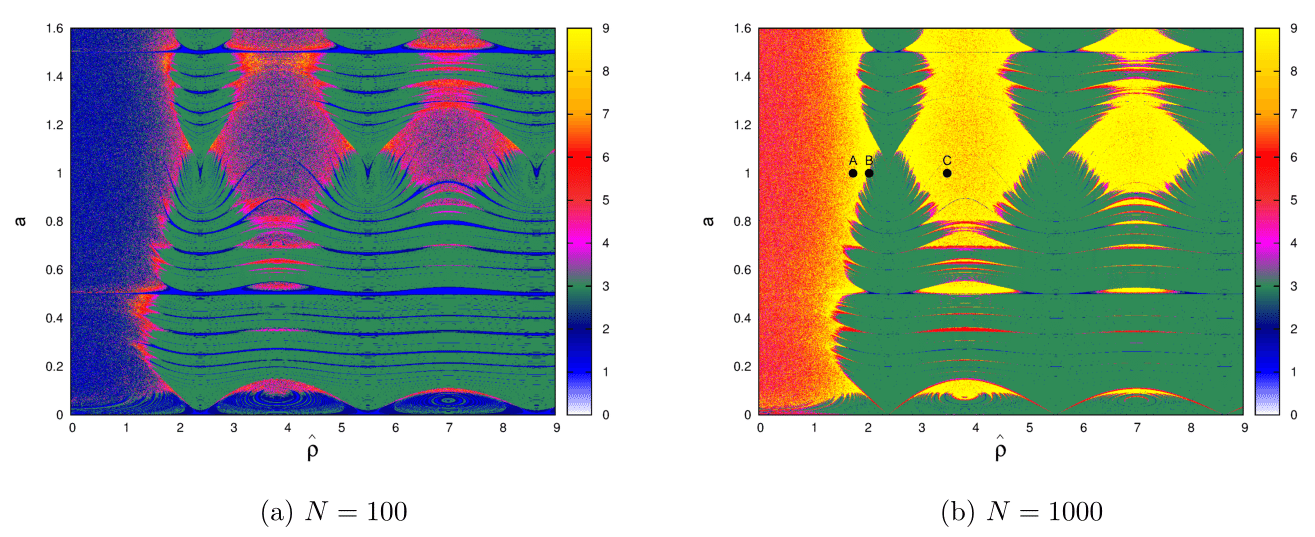}
\caption{GSNM breakup diagrams for $\hat{\rho}$ versus $a$, with $\bar{\rho}=0$ and $b=1.5$. Neighborhood size: $\epsilon=0.1$. 
Green points, corresponding to quasiperiodic dynamics, are concentrated near the zeros of $J_0$ and in domains with low $a$. (Color online)}
 \label{fig:aVSrhohatN100}
\end{center}
\end{figure}

\begin{figure}
\begin{center}
\includegraphics[scale=0.4]{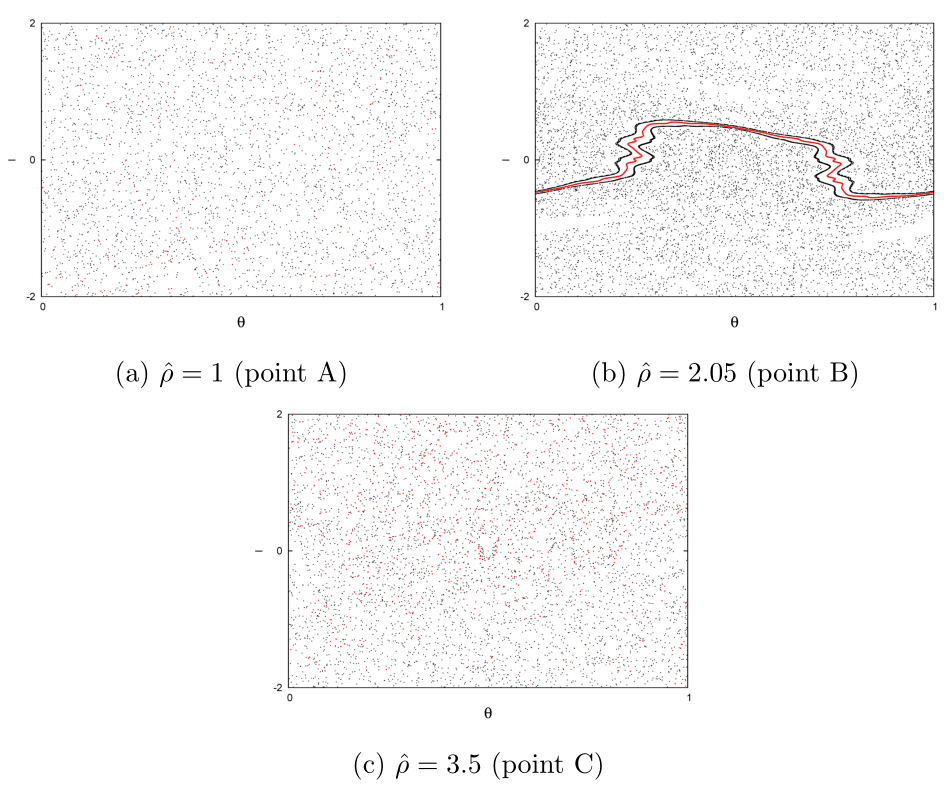}
\caption{Poincaré sections of the GSNM map for parameter values corresponding to 
points A, B and C, 
in Fig.~\ref{fig:bVSrhohatN100}, and $a=1$; $b=1.5$; $\bar{\rho}=0$. 
Consistent with Fig.~\ref{fig:bVSrhohatN100}-(b), the shearless curve is restored for 
$\hat{\rho}=2.05$. (Color online)} 
\label{fig:arhohat}
\end{center}
\end{figure}

Breakup diagrams with varying $a$ and $\hat{\rho}$ are shown
in Fig.~\ref{fig:aVSrhohatN100}. 
The number of iterations used to construct the first diagram was $N=100$. 
In the second diagram,  the number of iterations was increased to $N=1000$.
As before,  regions where $n_R=3$, which are indicators of non-chaotic dynamics,
are concentrated near the zeros of $J_0$. A high occurrence of points with $n_R=3$ is also observed in regions with low $a$.
When the number of iterations is increased from $N=100$ to $N=1000$, it s observed that the main geometric features remain the same, although several points with $n_R=1$ and $n_R=2$ transition to $n_R > 3$. 
Poincaré sections for parameters corresponding to points $A$, $B$, and $C$ in 
Fig.~\ref{fig:aVSrhohatN100}
are showed in Fig.~\ref{fig:arhohat}a-c. As expected, the shearless KAM
curve is broken at points $A$ and $C$, and restored at point $B$.

\begin{figure}
\begin{center}
\includegraphics[scale=0.35]{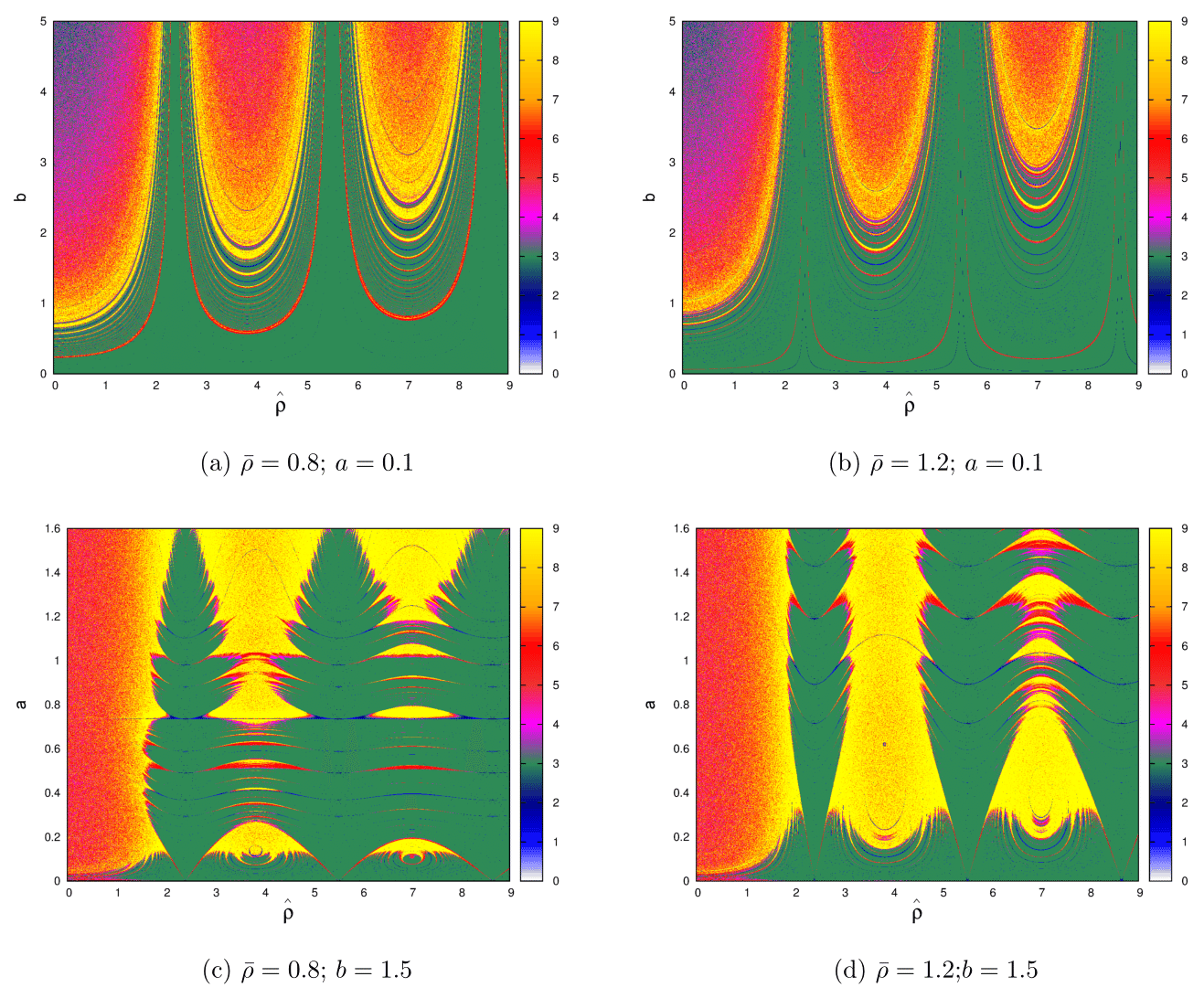}
\caption{Dependence of robustness of shearless curve on $\bar{\rho}$.  As shown in (a) and (b), the robustness of the shearless KAM increases when $\bar{\rho}$ increases with fixed $a=0.1$. On the other hand,  as shown in
(c) and (d), a different behavior is observed for fixed $b=1.5$. In all diagrams, $N=1000$ and $\epsilon=0.1$. (Color online)}
\label{fig:higherrhobar}
\end{center}
\end{figure}

Additional examples of breakup diagrams are presented in Figs.~\ref{fig:higherrhobar}~(a)-(d).
The parameters are the same as those in Figs.~\ref{fig:bVSrhohatN100}-(b) and \ref{fig:aVSrhohatN100}-(b), but the value of the fixed parameter $\bar{\rho}$ is higher. Comparing Figs.~\ref{fig:higherrhobar}~(a)-(b) to \ref{fig:bVSrhohatN100}-(b), 
it is observed that increasing $\bar{\rho}$    leads to an increase
in  the number of points with $n_R=3$ and to the robustness of the shearless KAM curve. 
As shown in Figs.~\ref{fig:higherrhobar}~(c)-(d), the distribution of points with $n_R=3$
changes when $\bar{\rho}$ increases, resulting in different critical thresholds but with no significant suppression of green points. 
Red and yellow points disappear in certain regions and reappear in others, as can be seen by comparing
Figs.~\ref{fig:aVSrhohatN100}-(b), \ref{fig:higherrhobar}-(c) and \ref{fig:higherrhobar}-(d).

Summarizing, in this section we used a computational technique based on the recurrence properties of the IPO 
to estimate the critical parameter values for the breakup the shearless KAM curve. In particular,
we computed breakup diagrams to understand the role of finite Larmor radius effects on the destruction and formation of shearless KAM curves.
It was observed that when $\hat{\rho}$ is close to a zero of $J_0$, the shearless curve becomes more robust to
perturbation in $b$. This is because, near the zeros of $J_0$, the GSNM's Hamiltonian is effectively  integrable. 
In particular, as $\hat{\rho}$ approaches a zero of $J_0$, 
the shearless curve, and the KAM curves that make up the nontwist transport barrier, are restored. The robustness of the shearless curve and the critical thresholds are also modified by increasing  $\bar{\rho}$.  

\section{Gyro-averaged quartic nontwist map}\label{sec:gqnm}

In this section, we propose another area-preserving map, the gyro-averaged quartic nontwist map (GQNM). A key property of the  
GQNM is that it exhibits a zonal flow bifurcation, similar to the one observed in the 
nontwist Hamiltonian system of Ref.~\cite{del-castillo-martinell2013}. The Hamiltonian system in Ref.~\cite{del-castillo-martinell2013}
is a drift-wave model of the ${\bf E}\times {\bf B}$ transport with FLR effects, and the zonal flow bifurcation corresponds to 
a bifurcation of the maximum of the ${\bf E}\times {\bf B}$ zonal flow velocity when the Larmor radius increases.

\begin{figure}
\begin{centering}
\includegraphics[scale=0.35]{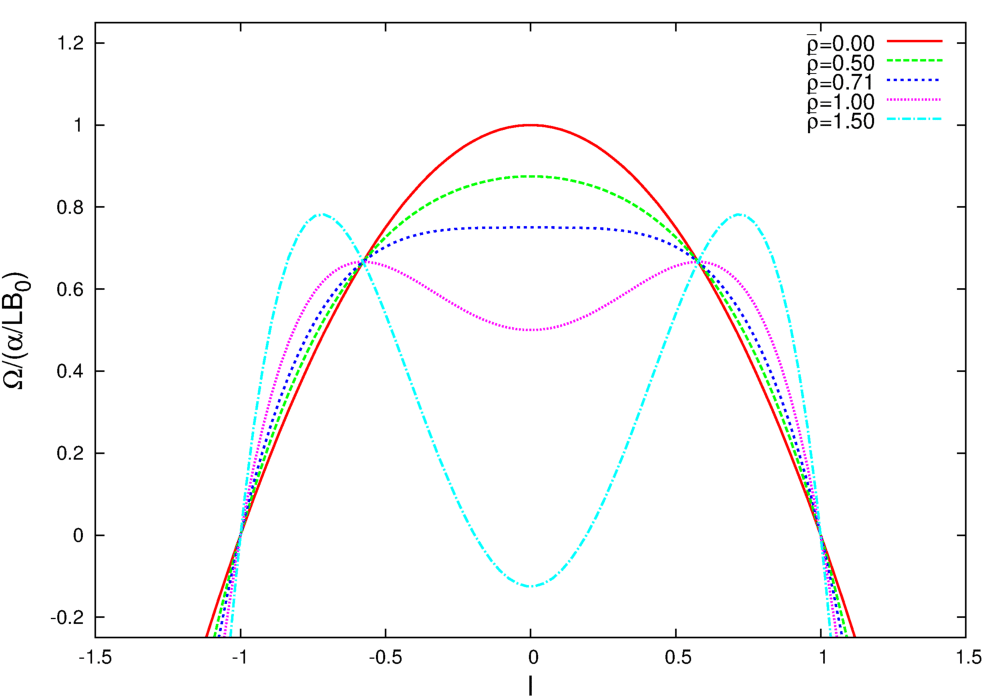}
\par
\end{centering}
\caption{Bifurcation of critical points in the frequency profile (\ref{eq:omegaaprox}) (Color online)}
\label{fig:omegaprof-gqnm}
\end{figure}

As in the previous maps, the GQNM is obtained from the gyro-averaged drift wave map in  Eqs.~(\ref{eq:dwmap-x})
and (\ref{eq:dwmap-y}). The equilibrium potential similar to the one proposed in Ref.~\cite{del-castillo-martinell2013}  
\begin{equation}
   \phi_{0}(x)=\alpha\tanh\left(\frac{x}{L}\right),\label{eq:tanh}
\end{equation}
where $\alpha$ and $L$ are dimensional constants. Applying the gyro-average operation (\ref{eq:FlrAverage}) to (\ref{eq:tanh})
gives
\begin{equation}
   \langle H_{0}(x)\rangle_{\varphi}=\frac{\alpha}{2\pi B_0}\int_{0}^{2\pi}\tanh\left(\frac{x}{L}+\frac{\rho}{L}\cos\varphi\right)\, d\varphi \, ,
\end{equation}
and from Eq.~(\ref{eq:freq}) we get 
the nonlinear frequency
\begin{equation}
   \Omega\left(I\right)=\frac{\alpha}{2\pi L B_0}\int_{0}^{2\pi} \sech^{2} \left(I+\bar{\rho}\cos\varphi\right)\, d\varphi \label{eq:omega-gqnm}
\end{equation}
where $I=x/L$, and $\bar{\rho}=\rho /L$. The zonal flow bifurcation results from a bifurcation of the critical point of the 
frequency profile. 
To analyze it in a simple setting, consider the Taylor series approximation of (\ref{eq:omega-gqnm})
for small values of $|I|$ and $\bar{\rho}$, 
\begin{equation}
   \Omega\left(I\right)\approx \frac{\alpha}{L B_0}\left(1-I^{2}\right)\left\{ 1-\bar{\rho}^{2}\left[\frac{3}{2}\left(1-I^{2}\right)-1\right]\right\}\, .\label{eq:omegaaprox}
\end{equation}
Figure \ref{fig:omegaprof-gqnm} shows the frequency profile in Eq.~(\ref{eq:omegaaprox}) for different values of $\bar{\rho}$. 
For $\bar{\rho}=0$, there is only one critical point, $\Omega'=0$, which corresponds to a maximum at $I=0$. 
Increasing $\bar{\rho}$ leads to an increase of the  profile's ``flatness'' and eventually to a transition in which the maximum 
bifurcates into a minimum and two maxima. The bifurcation threshold $\bar{\rho}_{b}$ can be determined
 from the condition,
\begin{equation}
 \frac{\partial^{2}\Omega}{\partial I^{2}}\bigg|_{I=0, \; \bar{\rho}=\bar{\rho}_{b}} = 0 \, ,\label{eq:threshold}
\end{equation}
which for Eq.~(\ref{eq:omegaaprox}) gives
\begin{equation}
   \bar{\rho}_{b}=1/\sqrt{2}. 
   \label{rhobarbif}
\end{equation}

Defining $I_n=x_n/L$, $\theta_n=ky_n/2\pi$, and $\hat{\rho}= k\rho$, and using Eq.~(\ref{eq:omegaaprox}), we get from 
Eqs.~(\ref{eq:dwmap-x}) and (\ref{eq:dwmap-y}), 
the GQNM 
\begin{align}
I_{n+1} =& I_{n}+bJ_{0}\left(\hat{\rho}\right)\sin\left(2\pi \theta_{n}\right)\label{eq:4dpolmap-I}\\
\theta_{n+1} =& \theta_{n}+a\left(1-I_{n+1}^{2}\right)\left\{ 1-\bar{\rho}^{2}\left[\frac{3}{2}\left(1-I_{n+1}^{2}\right)-1\right]\right\} ,\quad {\rm mod} \quad1\label{eq:4dpolmap-theta}
\end{align}
where $a=k \alpha /\omega_0 L B_{0}$, and the perturbation parameter is $b = 2\pi a A/\alpha \label{gqnm:b}$. As in the previous maps,  an effective perturbation parameter can be defined as $b_{ef} =bJ_{0}\left(\hat{\rho}\right)$. 
Like the GSNM, the GQNM reduces to the standard nontwist map for $\hat{\rho}=\bar{\rho}=0$.
In what follows, we discuss FLR effects on the GQNM's fixed points and nontwist transport barriers.

	\subsection{Fixed Points and nontwist transport barriers}

\begin{figure}[!h]
\begin{centering}
\includegraphics[scale=0.4]{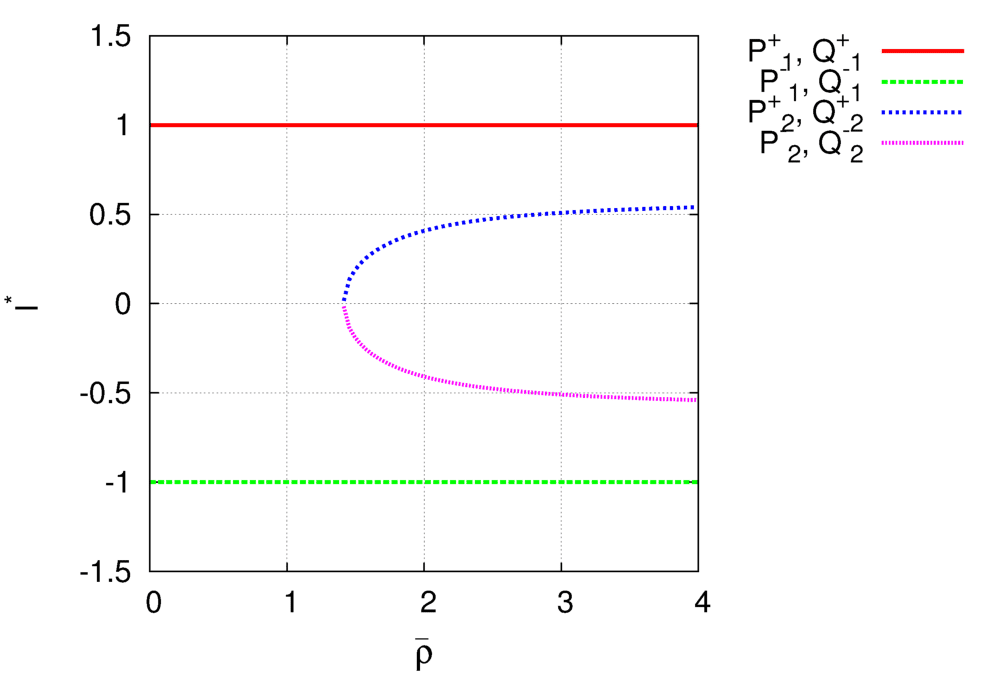}
\par\end{centering}
\caption{$I$ coordinates of fixed points with $m=0$ as function of $\hat{\rho}$. The set $\left\{P_{1}^{\pm}, Q_{1}^{\pm} \right\}$ 
exists for all values of $\bar{\rho}$. The set $\left\{P_{2}^{\pm}, Q_{2}^{\pm} \right\}$ 
only exists for $\bar{\rho} \geq 2\bar{\rho}_{b}=\sqrt{2}$. (Color online)}
\label{fig:gqnm-Icoord}
\end{figure}

\begin{figure}
\begin{centering}
\includegraphics[scale=0.4]{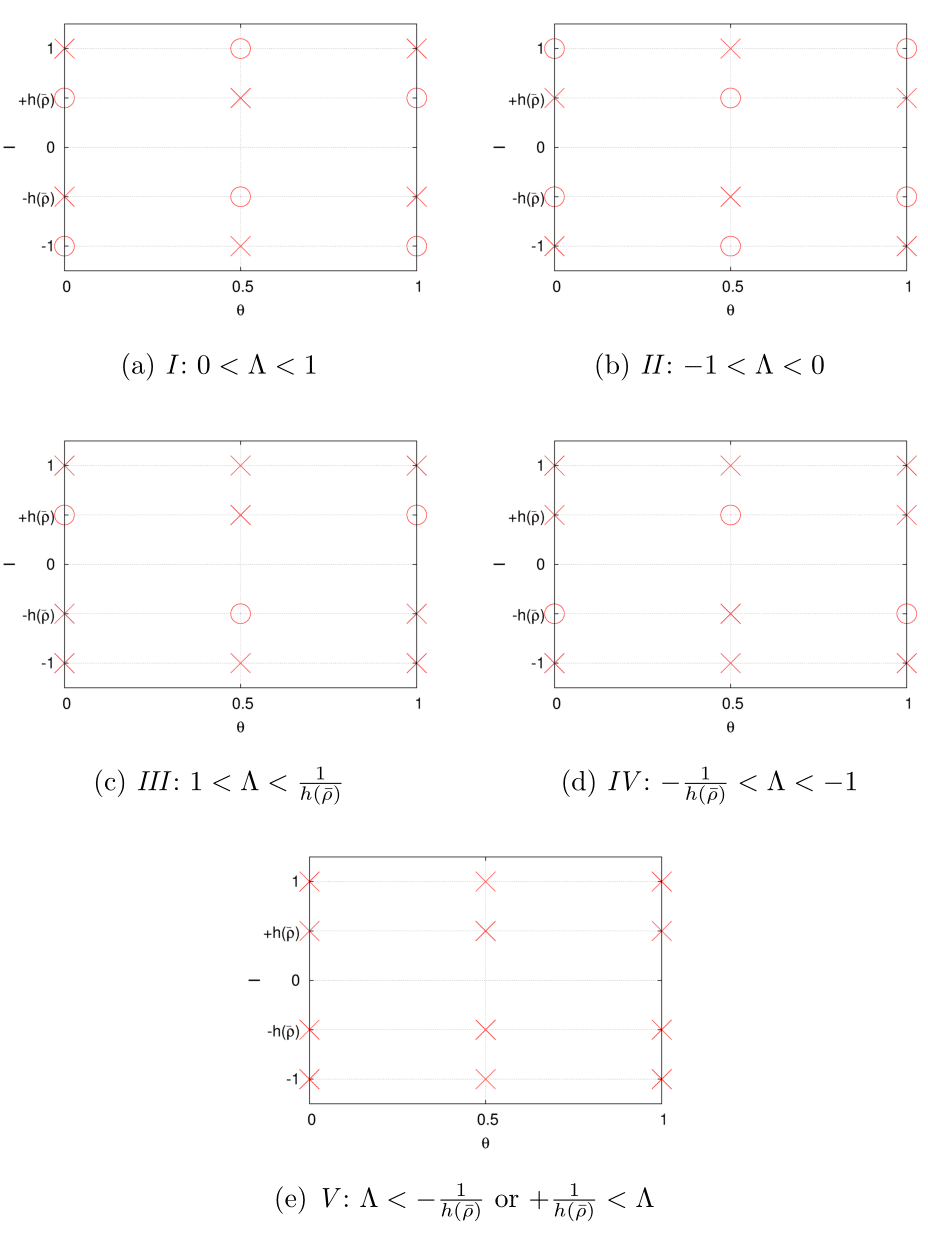}
\par
\end{centering}
\caption{Depending on the value of $\Lambda$, there are five possible configurations characterizing the stability of the period-one fixed points $\left\{P_{1}^{\pm}, Q_{1}^{\pm}, P_{2}^{\pm}, Q_{2}^{\pm} \right\}$ in the GQNM. (Color online)}
\label{gqnm-stabconfig}
\end{figure}

The fixed points $\left(\theta^{*}, I^{*}\right)$ of the GQNM satisfy
\begin{align}
0= & bJ_{0}\left(\hat{\rho}\right)\sin\left(2\pi\theta^{*}\right)\label{eq:4dpfixedpointper1th}\\
m= & a\left(1-I^{*2}\right)\left\{ 1-\bar{\rho}^{2}\left[\frac{3}{2}\left(1-I^{*2}\right)-1\right]\right\} ,\quad m\in\mathbb{\mathbb{Z}} \, .\label{eq:4dpfixedpointper1I}
\end{align}
For $m=0$, $a\neq0$, and $bJ_{0}\left(\hat{\rho}\right)\neq0$, there are four fixed points independently of the value of $\rho$, 
\begin{align}
   P_{1}^{\pm}  = \left(0,\pm1\right), \qquad  Q_{1}^{\pm}  = \left(\frac{1}{2},\pm1\right) \, .
\end{align}
If $\bar{\rho} \geq 2\bar{\rho}_{b} = \sqrt{2}$, there is an additional set of fixed points,
\begin{align}
   P_{2}^{\pm}  = \left(0,\pm h\left(\bar{\rho}\right)\right) \qquad Q_{2}^{\pm}  =\left(\frac{1}{2},\pm h\left(\bar{\rho}\right)\right) \, ,
\end{align}
where 
\begin{equation}
h\left(\bar{\rho}\right)=\sqrt{\frac{\bar{\rho}^{2}-2}{3\bar{\rho}^{2}}} \, .\label{eq:h}
\end{equation}
Figure \ref{fig:gqnm-Icoord} shows the $I$ coordinates of the fixed points with $m=0$.
$P_{1}^{\pm}$ and $Q_{1}^{\pm}$ exist for all $\bar{\rho} \geq 0$ and are always located at $I=\pm 1$.
$P_{2}^{\pm}$ and $Q_{2}^{\pm}$ only exist for $\bar{\rho} \geq 2\bar{\rho}_{b}=\sqrt{2}$ and their positions in the $I$ axis
are determined by the function $h\left(\bar{\rho}\right)$, which satisfies $0\leq h\left(\bar{\rho}\right) < \frac{1}{\sqrt{3}}$. 
The stability of the fixed points  $\left\{P_{1}^{\pm}, Q_{1}^{\pm}, P_{2}^{\pm}, Q_{2}^{\pm} \right\}$ can be analyzed by
evaluating the Green's residue in Eq.~(\ref{eq:res}) 
\begin{align}
R\left(P_{1}^{\pm}\right) & =-R\left(Q_{1}^{\pm}\right)=\mp\Lambda\left(a,b,\hat{\rho},\bar{\rho}\right) \label{eq:Rj1m0}\\
R\left(P_{2}^{\pm}\right) & =-R\left(Q_{2}^{\pm}\right)=\pm h\left(\bar{\rho}\right)\Lambda\left(a,b,\hat{\rho},\bar{\rho}\right)\label{eq:Rj-1m0}
\end{align}
where 
\begin{equation}
   \Lambda = \pi abJ_{0}\left(\hat{\rho}\right)\left(1+\bar{\rho}^{2}\right) \, .\label{gqnm-lambda} 
\end{equation}
As shown in Fig.~\ref{gqnm-stabconfig}, depending on the value of $\Lambda$, there are five possible stability configurations.
The number of elliptic fixed points reduces when $\Lambda$ increases, and for $\left|\Lambda\right|>\frac{1}{h\left(\bar{\rho}\right)}$
all the fixed points are hyperbolic. In general,
the first points to lose stability are the outer
ones. Note that as $\bar{\rho}$ approaches the zonal  bifurcation threshold, $1/h(\bar{\rho})$ diverges, and the range of the configurations III and IV increases. 

\begin{figure}
\begin{centering}
\includegraphics[scale=0.35]{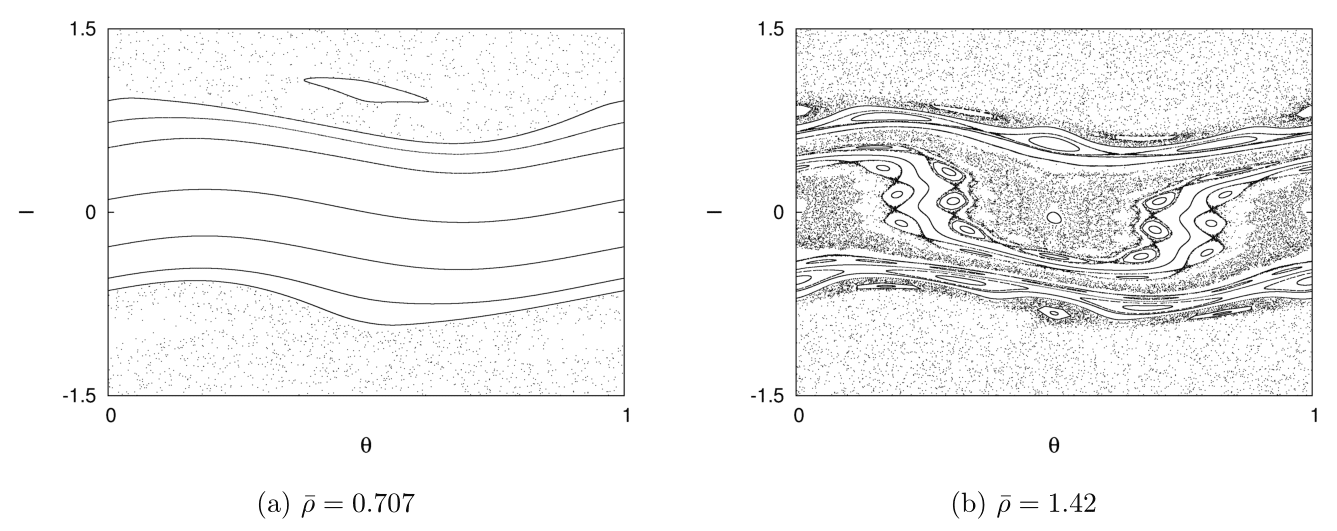}
\caption{(a) One NTB for $\bar{\rho} \lesssim \bar{\rho}_{b}$. (b) After the zonal flow bifurcation, three NTBs can be observed. Parameters: $a=0.5$; $b=1.35$; $\hat{\rho}=2.8$. }
 \label{fig:f1}
\end{centering}
\end{figure}

An important consequence of the zonal flow bifurcation is the occurrence of nontwist barrier bifurcations. In particular,
 for $\bar{\rho} \lesssim \bar{\rho}_{b}$, only one region with robust KAM curves is present 
 (Fig.~\ref{fig:f1})-(a), but
for $\bar{\rho} > \bar{\rho}_{b}$ (Fig. \ref{fig:f1})-(b) two additional NTBs appear. The outer NTBs are separated from the central one 
by two regions of confined chaotic orbits.

	\subsection{Robustness of the central shearless curve}

Our interest in studying the robustness of KAM curves as function
of $b$ resides on the fact that this parameter is proportional to the amplitude of the drift waves. For very small Larmor radius (or in the absence  of FLR corrections), high amplitude values can easily destroy all KAM curves.
However, this is  not always 
the case when FLR effects are taken into account. 
In particular, like in the GSNM (see Sec.~\ref{sec:ntb}),  the robustness of the  GQNM's shearless curve is significantly increased for $\hat{\rho}$ near a zero of $J_0$. 

\begin{figure}
\begin{centering}
\includegraphics[scale=0.35]{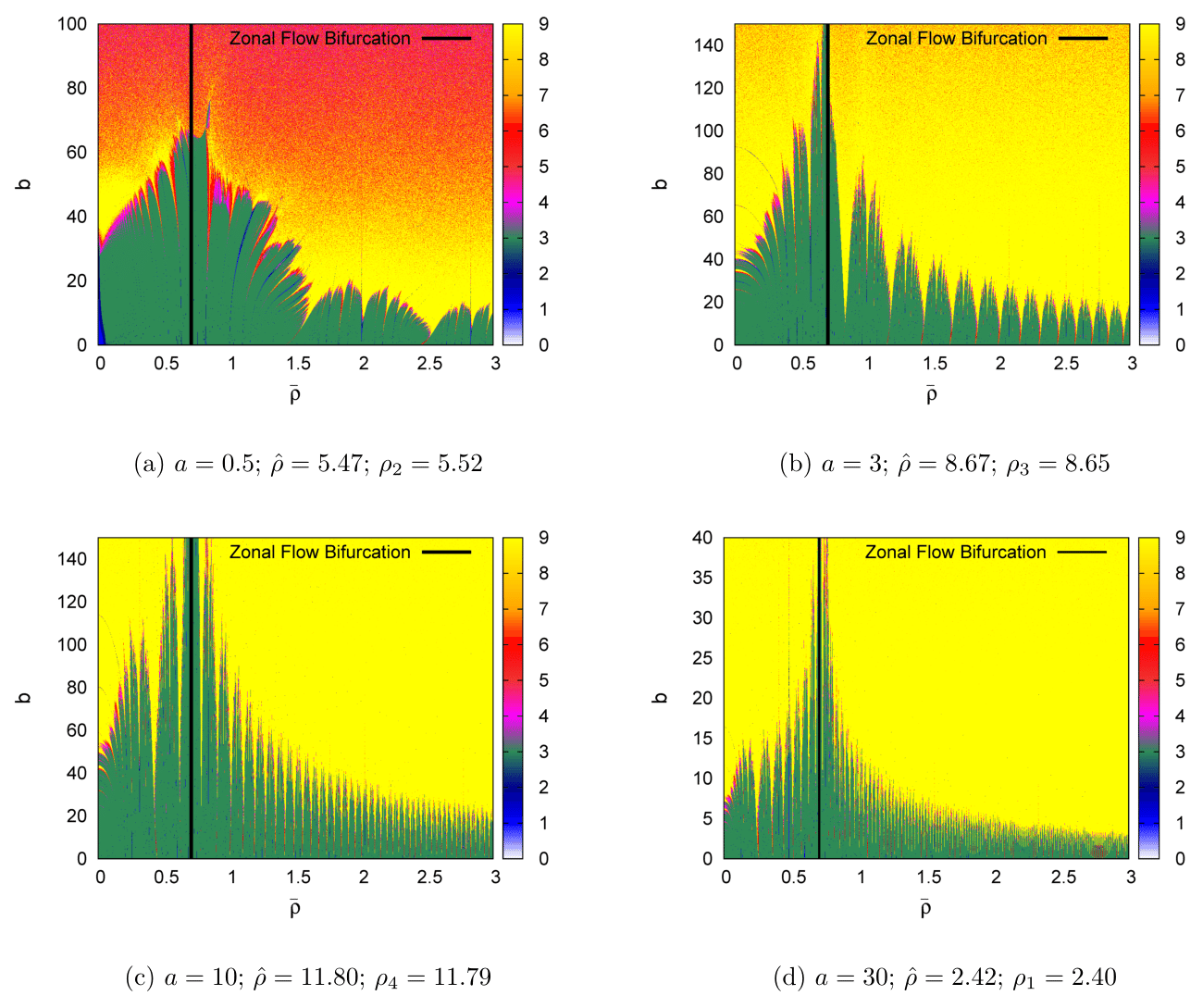}
\caption{For small $b_{ef}$ values, the robustness of the central shearless curve (and, thus, also of the central NTB) increases with 
the degree of flatness of the frequency's profile around the critical point.  The flatness can be controlled by the parameter $\bar{\rho}$ and becomes higher when $\bar{\rho}$
is close to the zonal flow bifurcation threshold $\bar{\rho}_{b}$ (shown with the solid black vertical line). The values of $b_{ef}$  are made small by setting $\hat{\rho}$ close to the zeros of $J_0$. The first four positive zeros are approximately given by: $\rho_1=2.40$; $\rho_2=5.52$; $\rho_3=8.65$; $\rho_4=11.79$. (Color online)}
\label{fig:CentraNTBRobustness}
\end{centering}
\end{figure}

Figures \ref{fig:CentraNTBRobustness}(a)-(d) show
breakup diagrams for the shearless KAM curve in the GQNM as function of $b$ and $\bar{\rho}$. In all cases, $b_{ef}$ is small, 
because the values of $\hat{\rho}$ are close to the zeros of $J_0$.
Under this condition,  and when  $\bar{\rho}$ is close to the zonal flow bifurcation threshold $\bar{\rho}_{b}$ in 
Eq.~(\ref{rhobarbif}), a significant increasing of the 
robustness of the shearless curve is  observed. For $\bar{\rho} < \bar{\rho}_{b}$, higher
values of $b$ are required for breakup as $\bar{\rho}$ approaches $\bar{\rho}_{b}$, and 
for $\bar{\rho} > \bar{\rho}_{b}$, the robustness is reduced as $\bar{\rho}$ increases.
That is, the robustness of the central NTB increases for $\hat{\rho}$ close to the zeros of $J_0$, and for  $\bar{\rho}$ close to $\bar{\rho}_{b}$.

It is interesting to note that, in the neighborhood  of the $I=0$ shearless point, the degree of ``flatness'' of the frequency's profile 
has a similar behavior. In particular, like the robustness of the central shearless curve, the  degree of ``flatness'' 
(measured by $\left|\Omega''(0)\right|$) is higher near the bifurcation threshold, but it becomes smaller elsewhere. 
These results provide support to the conclusion that the
robustness of NTBs in nontwist  maps depends on the flatness of the frequency profile at the maximum or minimum (critical points), and that, for small perturbation,
the robustness increases with the degree of flatness. 

The understanding of the role of the shear and the flatness of the frequency on the robustness of KAM curves is still an open problem. Among the first works addressing the role of shear is Ref.~\cite{finn_1975} where it was analytically and numerically shown that high shear reduces the strength of the perturbation required to break KAM surfaces. However this work limited attention to monotonic frequencies with non-vanishing shear (i.e., twist systems) and did not address the breakup of shearless KAM curves.  The robustness of NTBs  is also related to the strong KAM stability in nontwist maps \cite{Rypina}. According to the resonant overlap criterion \cite{chirikov79},  tori located between resonances break when the resonances overlap. Thus, one expects tori to be more resilient when the resonances' widths are small.
As shown in Ref.~\cite{Rypina}, in $3/2$-degrees-of-freedom nontwist Hamiltonian systems, the width, $\delta \Omega$, of second-order degenerate resonances scales as
\begin{equation}
   \delta \Omega \sim \left[\epsilon\left|\Omega''(I_0)\right|\right]^{2/3} \, ,\label{2orderreson}
\end{equation}
where $\epsilon$ is the amplitude of the perturbation, and $I_0$ is such that
\begin{align}
   \Omega'(I_0)=0 \label{omegader1} \qquad
   \Omega''(I_0)\neq0 \, ,
\end{align} 
which according to Eq.~(\ref{eq:twist-cond}) corresponds to the point where the twist condition is violated. 
When the system is slightly perturbed, second-order degenerate resonances appear in the neighborhood of the shearless curve, and their overlap leads to the break up of the  shearless curve.   

The central shearless curve of the GQNM is associated to
the critical point $I_0=0$ where the frequency in Eq.~(\ref{eq:omegaaprox}) satisfies (\ref{omegader1}) for $\bar{\rho}\neq\bar{\rho}_{b}$. Thus, in the
Hamiltonian system from which the GQNM is derived, second-order degenerate resonances might appear near the central shearless curve, and     
according to Eqs.~(\ref{2orderreson}) and (\ref{eq:threshold}), the width of these resonances scales as 
\begin{equation}
 \lim_{\bar{\rho} \to \bar{\rho}_{b}} \delta \Omega = 0 \, ,\label{limreswidth}
\end{equation}
because $\left|\Omega''(I_0)\right|$ vanishes for  $\bar{\rho}=\bar{\rho}_{b}$. Small values of $\left|\Omega''(I_0)\right|$ result in small degenerate resonance widths and a more
robust central shearless curve. As $\left|\Omega''(I_0)\right|$ is directly related to the flatness of the frequency's profile at $I_0$, 
Eq.~(\ref{2orderreson}) is consistent with the results Figs.~\ref{fig:CentraNTBRobustness}(a)-(b) and supports the conjecture relating the robustness of shearless curves to the flatness of the frequency 
at the degenerate points.  Similar arguments can be applied to interpret the results in Ref.~\cite{Firpo2009} where a study of the effect of  the flatness of the $q$-profile on the robustness of NTBs was presented. 

\section{Summary and Conclusions}
In this paper we studied  finite Larmor radius effects on discrete models of $\vec{E}\times\vec{B}$ chaotic  transport of test charged particles. 
The models are area-preserving Hamiltonian maps that assume an electrostatic potential consisting of the superposition of an equilibrium radial electrostatic field and a perturbation with a broad spectrum of drift waves. FLR effects were included by gyro-averaging the electrostatic potential and 
the maps were constructed from the integration of the resulting gyro-averaged $\vec{E}\times\vec{B}$ Hamiltonian system.

We constructed and studied three maps: the gyro-averaged standard map (GSM), the gyro-averaged standard nontwist map (GSNM), and the gyro-averaged quartic nontwist map (GQNM).  In the GSM, the frequency, i.e. the gyro-averaged $\vec{E}\times\vec{B}$  zonal flow velocity, has a monotonic radial profile and the map satisfies the twist condition. In the GSNM and GQNM cases  the frequency is non-monotonic and as a result the maps are  nontwist. Non-monotonic frequencies have degenerate points (i.e., 
points where the frequency is maximum or minimum), and in the vicinity of these points robust 
nontwist transport barriers (NTBs) tend to form. 
 
Typically, in area preserving maps, increasing the perturbation increases the chaos in the phase space. 
However, even for large perturbations, FLR effects can suppress chaos and restore broken KAM curves. 
The breakup diagrams provide the critical thresholds for 
the breakup of the shearless curve which is one of the robust KAM curves that constitute the NTB. 
To determine the breakup of the shearless curve we used an efficient and accurate  method based on the maximum number of different recurrence times. We showed that a relatively low number of iterations is sufficient
to determine the main features of the breakup diagrams. Although in general chaos is suppressed when 
the Larmor radius is  close to zeros of the zero-order Bessel function,  the breakup diagrams exhibit a highly nontrivial 
fractal-like dependence on the Larmor radius and the other parameters of the map. 

We also studied the role of the Larmor radius on the topology of the phase space 
and on bifurcations of the ${\bf E} \times {\bf B}$ zonal flow. 
In particular, we showed that the GSNM exhibits FLR dependent separatrix reconnection and 
obtained a formula determining the threshold for bifurcation from homoclinic to heteroclinic 
topologies as function of the Larmor radius. We also found that in the GQNM there is a critical value of the Larmor radius for which the maximum of the ${\bf E} \times {\bf B}$ zonal flow bifurcates into a minimum and two maxima.  These zonal flow bifurcations create additional shearless curves that play a highly nontrivial role on transport. 

Motivated by the key role that periodic orbits play on transport  we presented a detailed study of FLR effects on the 
location and stability of period-one orbits. 
In the GSNM there are four period-one orbits and, depending on the value of the Larmor radius, 
there are   three possible stability configurations. 
In the GQNM the situation is more complex due to the existence of zonal flow bifurcations. In particular, 
there can be up to eight period-one fixed points, four of which only exist for Larmor radii greater than a critical threshold. 
In this case, depending on the value of the Larmor radius, there are five stability configurations. 
It was found that the elliptic fixed points created by varying the Larmor radius are
strongly stable. The presence of elliptic points results in the formation of islands
that can trap orbits in the phase space. 

An important consequence of zonal flow bifurcations is the occurrence of up to three nontwist transport barriers. 
We focused on the central NTB and analyzed the robustness
of the shearless curve using breakup diagrams. The results showed that the robustness of the central shearless 
curve significantly increases when the Larmor radius is close to the critical threshold for the zonal flow bifurcation.
Based on this we argued that, in agreement with previous works,  the ``flatness'' of the frequency profile at the critical point is directly associated with the robustness of the central shearless curve. 

\section{Acknowledgments}
This work was made possible through financial support from the Brazilian research
agency FAPESP under grant 2013/00483-1.
The authors would like to acknowledge Celso V. Abud for valuable discussions.
DdCN was sponsored by the Office of Fusion Energy Sciences 
of the US Department of Energy at Oak Ridge National Laboratory, 
managed by UT-Battelle, LLC, for the U.S. Department  of Energy under contract DE-AC05-00OR22725.
JDF gratefully acknowledges the hospitality of the Oak Ridge National Laboratory where part of work was conducted.


\begin{thebibliography}{99}

\bibitem{kleva-drake-1984}
R.~G. Kleva, and J.~F. Drake,
Phys. Fluids {\bf 27}, 1686 (1984).

\bibitem{Pettini}
M. Pettini, A. Vulpiani, et al. Phys. Rev. A 38, 344 (1988)

\bibitem{ibere_horton_2008}
F.~A. Marcus, I.~L. Caldas, Z.~O Guimaraes Filho,  P.~J. Morrison, W. Horton, 
Y.~K Kuznetsov and
I.~L. Nascimento,  Phys. Plasmas {\bf 15},  112304 (2008). 

\bibitem{del-castillo-2000}
D. del-Castillo-Negrete, Phys. Plasmas {\bf 7},  1702 (2000).

\bibitem{Manfredi96} 
G. Manfredi and R. O. Dendy, Phys. Rev. Lett., {\bf 76}, 4360, (1996).

\bibitem{del-castillo-martinell2013}
J.J. Martinell and D. del-Castillo-Negrete, 
Phys. Plasmas {\bf 20}, 022303  (2013).

\bibitem{Annibaldi02}
S.V. Annibaldi, G. Manfredi, and R. O. Dendy, Phys. Plasmas {\bf 9}, 791 (2002).

\bibitem{Manfredi97}
G. Manfredi and R. Dendy, Phys. Plasmas {\bf 4}, 628 (1997).

\bibitem{gustafson} K.~ Gustafson, D. del-Castillo-Negrete and W. Dorland, Phys.\ Plasmas {\bf 16},
102309 (2008).

\bibitem{del-castillo-martinell2012}
D. del-Castillo-Negrete and J.J. Martinell,
Commun.\ Nonlinear Sci.\ Numer.\ Simulat.\ {\bf 17},  2031 (2012).

\bibitem{horton-1985}
W. Horton, Plasma Phys.\ Controlled Fusion
{\bf 27}, 9 (1985). 

\bibitem{Lee87}
W.~W. Lee, J. Comp. Phys {\bf 72}, 243 (1987).

\bibitem{Oda}
G. A. Oda and I. L. Caldas, Chaos, Solitons and Fractals, {\bf 5} 15 (1995).

\bibitem{Balescu98}
R. Balescu, Phys. Rev. E 58, 3 (1998).

\bibitem{del-castillo-1992}
D. del-Castillo-Negrete and P. J. Morrison, Bull. Am. Phys. Soc., Series II, 37, 1543 (1992)

\bibitem{Firpo2009}
L. Nasi and M.-C. Firpo, Plasma Physics and Controlled Fusion, {\bf 51}, 045006 (2009).


\bibitem{del-castillo-1993}
D. del-Castillo-Negrete and P. J. Morrison, Phys. Fluids A {\bf 5}, 948 (1993)

\bibitem{Beron-Vera10}
F. J. Beron-Vera, M. J. Olascoaga, M. G. Brown MG, et al. Chaos 20, 017514 (2010).

\bibitem{Budyansky09}
M. V. Budyansky, M. Yu. Uleysky, and S. V. Prants,  Phys. Rev. E, {\bf 79}, 056215 (2009). 

\bibitem{del-castillo-negrete-etal-1996}
D. del-Castillo-Negrete, J.~M. Greene, and P.~J.  Morrison,
Physica D {\bf 91}, 1 (1996).

\bibitem{Rypina}
I.I Rypina, M. G. Brown, et al.  Phys. Rev. Lett. {\bf 98}, 104102 (2007).


\bibitem{Portela07}
J. S. E. Portela, I. L. Caldas et al. International Journal of Bifurcation and Chaos in Applied Sciences and Engineering, {\bf 17}, 1589, (2007).

\bibitem{Szezech09}
J. D. Szezech, I. L. Caldas, et al. Chaos: An Interdisciplinary Journal of Nonlinear Science, {\bf 19}, 043108 (2009).

\bibitem{Caldas}
I.L. Caldas,  R.L. Viana et al. Commun.\ Nonlinear Sci.\ Numer.\ Simulat.\ {\bf 17}, 2021 (2011)

\bibitem{de-la-Llave-2000}
A. Delshams and R. de-la-Llave, SIAM J. Math. Anal., 31, 1235 (2000)

\bibitem{Altmann}
E. G. Altmann, G. Cristadoro, and D. Pazó, Phys. Rev. E, {\bf 73}, 056201 (2006).

\bibitem{Zou}
Y. Zou, D. Pazó, et al. Phys. Rev. E, {\bf 76}, 016210 (2007).

\bibitem{Abud}
C.V. Abud, PhD Thesis, University of São Paulo, (2013).

\bibitem{Nicholson}
D.R Nicholson, ``Introduction to Plasma Theory''. John Wiley and Sons, Inc. New York, 1983.

\bibitem{horton-1998}
W. Horton, H.-B. Park, J.-M. Kwon, D. Strozzi, P. J. Morrison, and D.-I. Choi, Phys. Plasmas {\bf 5}, 11 (1998).

\bibitem{chirikov79}
B. V. Chirikov, Phys. Rep., {\bf 52}, 263, (1979).

\bibitem{taylor}
J. B. Taylor, Culham Lab. Prog. Report CLM-PR-12, (1969).

\bibitem{Meiss92}
J.D. Meiss, Rev. Mod. Phys. {\bf 64}, 3 (1992).

\bibitem{Greene79}
J.M Greene, J. Math. Phys. {\bf 20}, 1183 (1979)

\bibitem{shinohara}
S. Shinohara and Y. Aizawa, Prog. Theor. Phys. {\bf 97}, 379 (1997).

\bibitem{Petrisor}
E. Petrisor, Int. J. Bifurcation Chaos, 11, 497 (2001)

\bibitem{Wurm}
A. Wurm, A. Apte,  et al.  CHAOS {\bf 15}, 2023108 (2005).

\bibitem{Slater}
N. Slater, Proc. Cambridge Philos. Soc. {\bf 63}, 1115 (1967).

\bibitem{finn_1975}
J.~M. Finn, Nuclear Fusion, {\bf 5}, 845 (1975). 

\end{thebibliography}
\end{document}